\documentclass[fleqn,usenatbib]{mnras}

\usepackage{newtxtext,newtxmath}

\usepackage[T1]{fontenc}

\DeclareRobustCommand{\VAN}[3]{#2}
\let\VANthebibliography\thebibliography
\def\thebibliography{\DeclareRobustCommand{\VAN}[3]{##3}\VANthebibliography}

\usepackage{graphicx}	
\usepackage{amsmath}	
\usepackage{xcolor}

\newcommand{\sout}[1]{\bgroup 
 \markoverwith
 {\textcolor{red}{\rule[0.5ex]{5pt}{1pt}}}\ULon{#1}}

\newcommand{\pkdgrav}{{\sc pkdgrav}}
\newcommand{\gala}{{\sc Gala}~}

\newcommand{\kms}{\mbox{km s$^{-1}~$}} 
\newcommand{\kmse}{\mbox{km s$^{-1}$}}

\newcommand{\dgr}{$^{\circ}~$}
\newcommand{\dgre}{$^{\circ}$}

\newcommand{\hi}{H{\footnotesize I} }

\newcommand{\meanz}{$\left<z\right>$}

\title[Modeling recent MC interactions]{Modeling the recent interactions between the Magellanic Clouds and Milky Way}

\author[B. R. Garver et al.]{
Bethany R. Garver,$^{1}$\thanks{E-mail: bethanygarver@montana.edu}
David L. Nidever,$^{1,2}$
Victor P. Debattista,$^{3}$\thanks{E-mail: vpdebattista@lancashire.ac.uk}
Nathan Deg$^{4}$
\\
$^{1}$Department of Physics, Montana State University, P.O. Box 173840, Bozeman, MT 59717-3840\\
$^{2}$Center for Computational Astrophysics, Flatiron Institute, 162 Fifth Avenue, New York, NY 10010, USA\\
$^{3}$Jeremiah  Horrocks  Institute, University of Lancashire, Preston, PR1 2HE, UK\\
$^{4}$Department of Physics, Engineering Physics, and Astronomy, Queen’s University, Kingston ON K7L 3N6, Canada\\
}

\date{Accepted XXX. Received YYY; in original form ZZZ}

\pubyear{2026}

\begin{document}
\label{firstpage}
\pagerange{\pageref{firstpage}--\pageref{lastpage}}
\maketitle

\begin{abstract}
The Large and Small Magellanic Clouds (LMC and SMC, respectively) are the largest satellite galaxies of the Milky Way (MW) and their interactions with each other have given rise to multiple stellar substructures in their periphery as well as the gaseous Magellanic Stream. 
To better understand the origin of the stellar substructures and constrain their past orbit, we model the past 2.5 Gyr of the interactions between the MW and the LMC and SMC using N-body simulations. Due to the strong interactions, analytical orbit integrations are insufficient to analyze the past galaxy orbits accurately.
Therefore, we use a genetic algorithm in combination with N-body simulations to determine the LMC and SMC initial positions and velocities 2.5 Gyr ago that result in the Magellanic Clouds (MCs) arriving near their observed locations and velocities at the current time. After running $\sim$8,000 simulations, our best matching model includes two close interactions between the MCs (940 Myr and 140 Myr ago) and reproduces some observed features of the MCs, including the LMC disc warp, a ring-shaped overdensity in the LMC, the tidal expansion of the SMC, and a greater distance dispersion on the eastern side of the SMC. The LMC disc warp is caused by the most recent interaction with the SMC, which occurred $\sim$140 Myr before the present. The interaction causes global ripples in the LMC disc with a mean amplitude of 1.3 kpc.
\end{abstract}




\section{Introduction}

The Large Magellanic Cloud (LMC) and Small Magellanic Cloud (SMC) are nearby galaxies interacting with each other and with the Milky Way (MW).  Their structure has been studied for decades with optical and near-infrared (NIR) photometry \citep[e.g.,][]{vandermarel2001,Mackey2016,Nidever2017}, optical/NIR spectroscopy \citep[e.g.,][]{VanderMarel2002,Carrera2008,Nidever2020}, UV \citep[][]{Hagen2017}, H$\alpha$ \citep[][]{Barger2013}, radio continuum \citep[e.g.,][]{Klein1993,Filipovic1998}, molecular gas \citep[e.g.,][]{Wong2011}, 21cm \hi \citep[e.g.,][]{Putman2003,Nidever2008}, and more.


Recent wide-field surveys and spectroscopic studies have revealed that the Magellanic Clouds host a wealth of substructure, providing strong evidence of their dynamic interaction history. \citet{Majewski2009} and \citet{Nidever2011} used pencil-beam Washington+{\em DDO}51 photometry with follow-up spectroscopy of giant stars to discover evidence for very extended stellar components around the MCs (LMC: $R$$\approx$20\degr, SMC: $R$$\approx$12\degr).  \citet{Saha2010} used deep photometry to track main-sequence turnoff (MSTO) stars out to a radius of $\sim$16\dgr to the north.
\citet{Mackey2016} first mapped the northern stream of the Large Magellanic Cloud (LMC), and subsequent work with deeper imaging revealed that this structure extends even farther \citep{Belokurov2019} and includes two distinctive southern `hook'-shaped features \citep{Mackey2018}. Using red clump stars from the Survey of the Magellanic Stellar History \citep[SMASH;][]{Nidever2017, Nidever2021}, \citet{Choi2018a, Choi2018b} mapped the three-dimensional geometry of the LMC disc and uncovered a pronounced southern warp extending roughly 4 kpc below the disc plane in the direction of the Small Magellanic Cloud (SMC). \citet{Saroon2022} later identified a northern warp with {\it Gaia} DR2 data that bends in the same direction, giving the LMC disc a `U'-shaped morphology—an unexpected feature if the distortion were purely tidal in origin. In addition, the outer stellar density profile of the LMC shows significant irregularities. \citet{Nidever2019a} found that while the overall profile can be described by a double-exponential or exponential-plus-power-law form, there is considerable scatter about the best-fitting model, indicating substantial substructure in the outskirts. Kinematically, \citet{Olsen2011} uncovered a population of metal-poor, counter-rotating stars in the LMC, providing further evidence for past interactions or accretion events. 
Finally, \citet{Nidever2020} used SDSS-IV / APOGEE-2S \citep{Majewski2017} high-resolution spectroscopy to map the chemical enrichment of both Clouds, finding similarly low [$\alpha$/Fe] `knees' at [Fe/H] $\approx$ $-$2.2 and evidence for recent starburst activity, further linking the Clouds’ chemical evolution to their dynamical interaction history.

\begin{figure}
    \centering
    \includegraphics[width=0.95\linewidth]{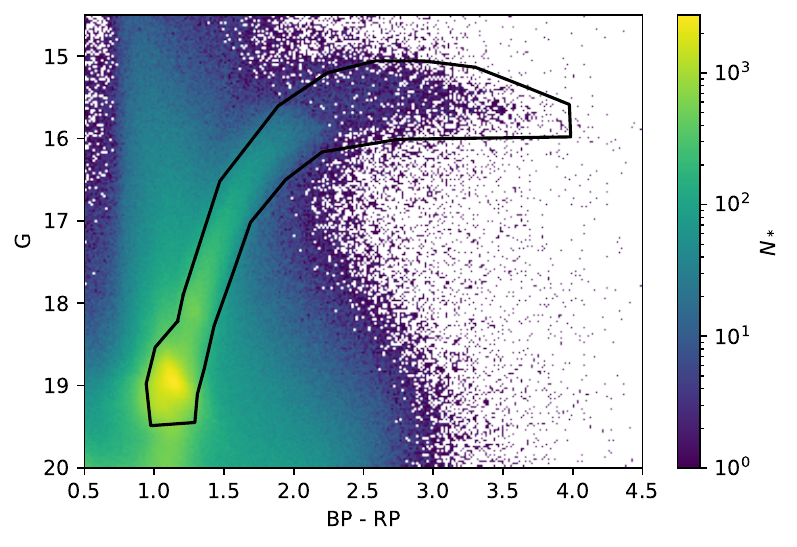}
    \caption{Color magnitude diagram of {\it Gaia} EDR3 MC stars. The black boundary shows the cuts we use for MC red clump and red giant branch stars.}
    \label{fig:cmdcuts}
\end{figure}

The SMC exhibits a similarly complex structure, with both its morphology and kinematics bearing clear signatures of tidal influence from the LMC. Early photographic studies revealed a distance bimodality in the eastern SMC \citep{Hatzidimitriou1989}, later confirmed with deep photometric data by \citet{Nidever2013}, who found that the more distant eastern component resembles the central and western regions of the SMC, while the nearer eastern component consists of tidally stripped material. Spectroscopic evidence supports this picture with \citet{Almeida2024} showing that the distant eastern stars have a metallicity distribution function (MDF) similar to that of the western population, whereas the near eastern stars are more metal-rich, resembling the central SMC population—consistent with their origin as recently stripped stars. The SMC also possesses extended stellar features, including the SMC Northern Overdensity (SMCNOD), discovered by \citet{Pieres2017} using DES data \citep{DES2016}, located roughly 8\dgr from the SMC center. The timing and geometry of these features provide important constraints on the most recent close encounter between the Clouds. \citet{Choi2022} used {\it Gaia} measurements of LMC disc distortions to infer the timing and impact parameter of this interaction.  Observational constraints from {\it Gaia} DR2, including proper motions and radial velocities of red giant stars, were used by \citet{Zivick2021} to model SMC kinematics, revealing a modest intrinsic rotation of $\sim$10–20 \kms and a tidal expansion component of $\sim$10 \kms~kpc$^{-1}$.
\citet{Massana2024} used {\it Gaia} XP stellar parameters from \citet{Andrae2023} to map metallicities across the MCs and found that both the SMCNOD and its southern counterpart, the SMC Southern Overdensity (SMCSOD), are metal-poor with MDFs similar to those in the SMC outskirts.
Together, these discoveries paint a picture of a dynamically active Magellanic system, where repeated interactions between the LMC and SMC have produced tidal warps, streams, and extended stellar envelopes.


Simulations of the Magellanic Clouds have progressed from early low-resolution N-body approaches \citep[e.g.,][]{Lin1977,Kunkel1979,Murai1980} to modern hydrodynamical models that incorporate detailed Milky Way interactions and internal cloud dynamics. \citet{Ruzicka2010} performed a broad parameter-space search using a genetic algorithm \citep{Wahde1998,Theis1999} with test-particle simulations, providing some initial constraints on the orbital and interaction history of the Clouds. \citet{Connors2006} N-body simulations reproduced the Leading Arm and the two filaments of the trailing Stream using tidal interactions between the MCs and MW. After \citet{Besla2007} discovered that the MCs are very likely on their first infall into the MW, a newer generation of simulations by \citet{Besla2012} and \citet{Diaz2012} explored the interactions between the LMC and SMC and were able to reproduce key features of the Magellanic Stream, including its bifurcated filamentary structure and predicted the SMC `counterbridge'. 

Hydrodynamical simulations have further illuminated the role of gas physics in shaping the Magellanic system. \citet{Lucchini2020} included a hot Magellanic corona in their models, which shields leading gas from ram pressure stripping by the MW halo and naturally produces the observed Leading Arm Feature \citep[LAF;][]{Putman1998,Bruens2005}. While \citet{Lucchini2021} suggested that the trailing Magellanic Stream could lie as close as $\sim$20 kpc, recent stellar detections within the Stream at $\sim$100 kpc \citep{Zaritsky2020,Chandra2023} indicate a greater distance. Supporting these models, \citet{PriceWhelan2019} and \citet{Nidever2019b} identified a young (117 Myr) star cluster in the MW formed from Leading Arm gas.

\begin{figure}
    \centering
    \includegraphics[width=0.95\linewidth]{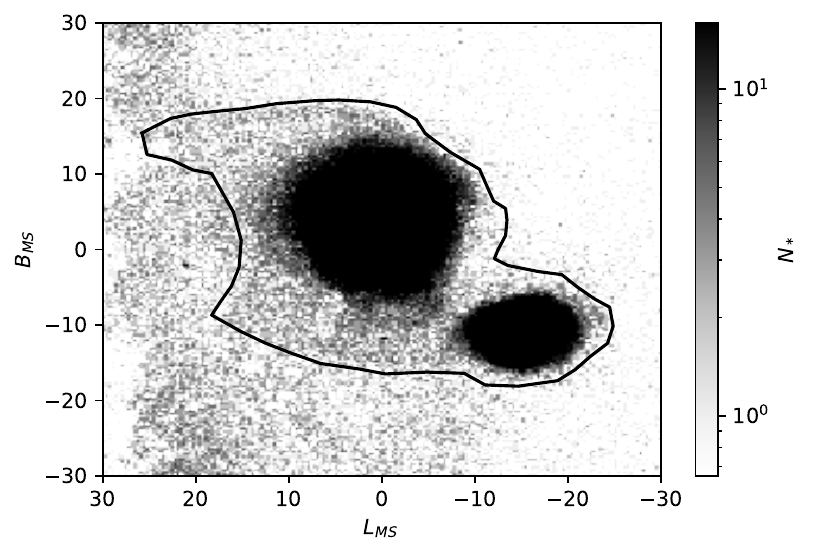}
    \caption{Density map of {\it Gaia} MC giant stars. The black line shows the selection we used on the map of the stars on Magellenic Stream coordinates $L_{MS}$ and $B_{MS}$.}
    \label{fig:mapcuts}
\end{figure}

Comprehensive simulation suites such as KRATOS \citep{JimenezArranz2024}, and the studies by \citet{Sheng2024} and \citet{Lucchini2024} explore a wide range of parameters including halo masses, corona properties, and orbital histories, providing frameworks to reproduce both the large-scale gaseous features and the complex stellar substructures observed today. Additionally, \citet{Vasiliev2024} demonstrated methods for evolving an LMC within a Milky Way potential while recovering key kinematic and structural properties, offering a practical approach to connecting simulations with observations. However, reliably reconstructing the orbital history of the MCs that models their interactions accurately and reproduces their current positions, velocities and substructures remains challenging and elusive. In particular, their close interactions are challenging to model analytically because of the strong dynamical friction and tidal distortions.

Here, we attempt to tackle the challenge of recovering the complex recent interaction history of the MCs as well as their stellar substructures with a large suite of N-body simulations.  In Section \ref{sec:data}, we discuss the {\it Gaia} and APOGEE datasets used for observational comparison.  The simulations are described in Section \ref{sec:simulations} and details of setting up the interactions and genetic algorithm are provided in Section \ref{sec:methods}.
A detailed comparison of our simulations with observed features is given in Section \ref{sec:results}.
Finally, we discuss the implications of our work in Section \ref{sec:discussion} and provide our conclusions in Section \ref{sec:conclusions}.

\begin{figure*}
    \centering
    \includegraphics[width=1.00\textwidth]{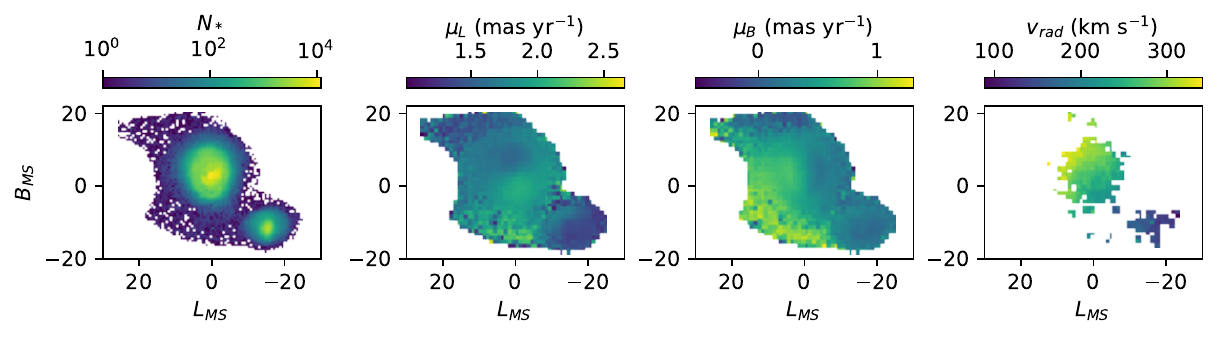}
    \caption{Maps of the observed {\it Gaia} and SDSS MC data. {\em left:} Number of stars, {\em center left:} mean proper motion in the $L_{\rm MS}$ direction, {\em center right:} mean proper motion in the $B_{\rm MS}$ direction, and {\em right} mean line-of-sight velocity.}
    \label{fig:dataplots}
\end{figure*}

\section{Data}
\label{sec:data}

The observational data we use to generate the MC proper motion and density maps come from {\it Gaia} EDR3 \citep{Gaia2021}, while the radial velocity measurements come from SDSS-IV / APOGEE DR17 \citep{SDSS2022}. To filter the {\it Gaia} data, we use a process similar to what is described by \citet{Belokurov2019}. A CMD filter is applied to select MC red clump (RC) and red giant branch (RGB) stars, as shown in \autoref{fig:cmdcuts}. In addition, stars with proper motions outside the range of $0.9 \leq \mu_{L_{\rm MS}} \leq 2.8$ mas yr$^{-1}$ and $-0.6 \leq \mu_{B_{\rm MS}} \leq 1.4$ mas yr$^{-1}$ are removed, where $\mu_{L_{\rm MS}}$ and $\mu_{B_{\rm MS}}$ are proper motions in the Magellanic Stream (MS) coordinate system \citep{Nidever2008}. Foreground stars are also removed with a parallax cut $\varpi > 0.2$. Finally, we select stars in $L_{\rm MS}$--$B_{\rm MS}$ space to include the main stellar features of the interaction between the MCs (see \autoref{fig:mapcuts}).
The observational maps of density, proper motions, and radial velocities are shown in \autoref{fig:dataplots}.

\section{Simulations}
\label{sec:simulations}

We model the interactions between the MW and MCs using N-body simulations evolved with the \pkdgrav~software package \citep{Stadel2001}. \pkdgrav~is a highly efficient, fully parallel N-body code that employs adaptive spatial and temporal resolution as well as a k-D tree–based force algorithm to enable large cosmological simulations for studying galaxy formation. Low-resolution simulations are used with the genetic algorithm for improved speed. Subsequently, a variety of higher-resolution simulations are performed for the best-fitting orbit.

For the low-resolution N-body simulations used with the genetic algorithm, the three galaxies are initialized as follows: The LMC is modeled with $1.62\times 10^5$ star particles and $2.38\times 10^5$ dark matter particles.
The assumed stellar mass of the LMC is $7.20 \times 10^9~M_{\odot}$ and the dark matter halo has a mass of $1.76 \times 10^{11} M_{\odot}$ \citep{Erkal2019b}. The stellar mass is higher than the observed value of $2.7\times 10^9 M_{\odot}$ \citep{Besla2015} because it includes the gas mass as well as mass that has been stripped from the LMC in the past few Gyr. The SMC has $2.25 \times 10^4$ star particles and $2.9\times 10^4$ dark matter particles. Its stellar mass is $1.06 \times 10^9~M_{\odot}$ and its dark matter halo is $2.02 \times 10^{10} M_{\odot}$ \citep{Besla2012}. The MW has $2.4\times 10^4$ star particles and $2.5\times 10^4$ dark matter particles. Its stellar mass is $5.25 \times 10^{10}~M_{\odot}$ and its dark matter halo has mass $1.16 \times 10^{12} M_{\odot}$. These three models are each evolved in isolation for 6 Gyr. The `evolved' galaxies are then used as inputs for the genetic algorithm simulations where all three combined are run for another 2.5 Gyr.

After running the genetic  algorithm, we use $10\times$ higher resolution LMC, SMC, and MW models. The higher resolution makes it easier to find features in the MCs while having minimal effect on the orbits. In simulations comparing the orbits of the LMC with the lower and higher resolution MWs, the final LMC locations are less than 2 kpc apart. 

In addition to our fiducial high-resolution LMC model (model A), we run the simulation with the same initial positions and velocities for five other LMC models: B, which has a thicker disc, C, which has a thinner disc, D, which has a hotter disc, F, which has a more compact dark matter halo, and G, which has a more diffuse dark matter halo. The properties of these models and the SMC model are given in \autoref{tab:lmcmodels}.

\begin{table*}
    \centering
    \begin{tabular}{cccccccc}
        \hline
        Model & $M_h$ & $R_h$ & $\sigma_h$ & $M_d$ & $R_d$ & $Z_d$ & $\sigma_R$\\
         & ($10^{11}~\rm{M}_{\odot}$) & (kpc) & ($\kms$) & ($10^{9}~\rm{M}_{\odot}$)  & (kpc) & (kpc) & ($\kms$)\\
         \hline
        LMC A & 1.76 & 21.4 & 222 & 7.20 & 2 & 0.28 & 45\\
        LMC B & 1.75 & 21.4 & 222 & 7.20 & 2 & 0.41 & 45\\
        LMC C & 1.77 & 21.4 & 222 & 7.20 & 2 & 0.22 & 45\\
        LMC D & 1.76 & 21.4 & 222 & 7.20 & 2 & 0.28 & 56.25\\
        LMC F & 1.76 & 13 & 237 & 7.20 & 2 & 0.28 & 45\\
        LMC G & 1.76 & 32 & 216 & 7.20 & 2 & 0.28 & 45\\
        SMC & 0.202 & 7.3 & 125 & 1.06 & 1.4 & 0.26 & 20\\
         \hline
    \end{tabular}
    \caption{Properties of the different LMC models and the SMC model we use.}
    \label{tab:lmcmodels}
\end{table*}

\begin{table}
    \centering
    \begin{tabular}{ccc}
    \hline
         Galaxy&  Observed Position& Observed Velocity\\
         & ($x$, $y$, $z$) [kpc] & ($v_x$, $v_y$, $v_z$) [km s$^{-1}$]\\
         \hline
         LMC&  ($-$1.0, $-$40.9, $-$27.7)& ($-$57$\pm$13, $-$226$\pm$15, 221$\pm$19)\\
         SMC&  (14.9, $-$38.1, $-$44.2)& (18$\pm$6, $-$179$\pm$16, 174$\pm$13)\\
         \hline
    \end{tabular}
    \caption{The observed positions and velocities that we compare to the simulations.}
    \label{tab:obs}
\end{table}

\section{Methods}
\label{sec:methods}

We run simulations with just the LMC and SMC, as well as simulations that include both MCs and the MW. For both groups of simulations, we save snapshots every 20 Myr and calculate the median particle position and velocity for the LMC and SMC. The median is used rather than the center of mass because it is less sensitive to outliers. We compare these values to the observed center of mass (COM) positions and velocities shown in \autoref{tab:obs}. The LMC velocity comes from \cite{Kallivayalil2013} and the SMC velocity comes from \cite{Zivick2018}.

In both groups of simulations, we use a genetic algorithm in which we vary the initial positions, velocities, and orientations of the simulated MCs and compare them to the observed COM positions and velocities, as well as density, proper motions in the $L_{MS}$ and $B_{MS}$ directions, and radial velocities at the `current' time.  The current time is defined when the LMC center reaches $L_{MS}$=0\degr, with the Sun at the position of (X,Y,Z)=($-$8.0 kpc, 0.0 kpc, 0.0 kpc) in the MW-centric frame.

\begin{figure}
    \centering
    \includegraphics[width=0.95\linewidth]{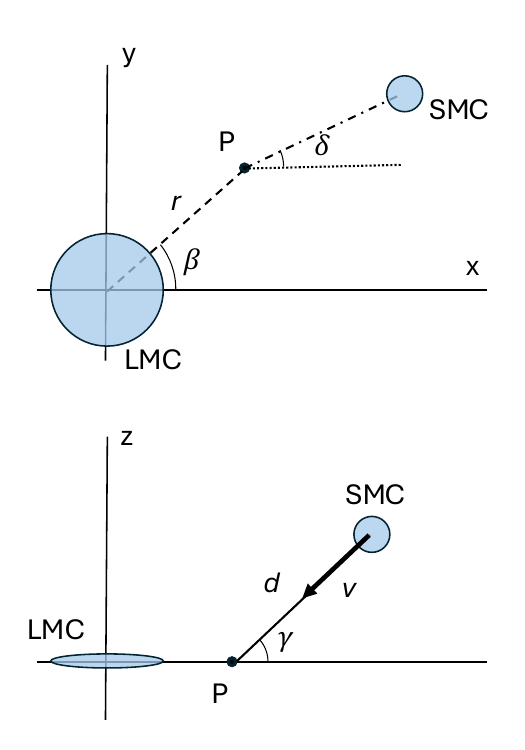}
    \caption{Diagram of the coordinates we define for our two-galaxy simulations. {\em Top:} projection onto the $x-y$ plane. {\em Bottom:} projection onto the plane perpendicular to the $x-y$ plane that includes the initial location of the SMC center and point P.}
    \label{fig:twogalaxycoords}
\end{figure}

\subsection{Genetic Algorithm}
\label{subsec:genetic}

We employ a genetic algorithm to thoroughly and efficiently explore the 9-dimensional parameter-space of initial conditions that produces the best-fitting final LMC/SMC positions and velocities. A genetic algorithm is a stochastic optimization technique that explores parameter space by mimicking natural selection, evolving a population of candidate solutions through operations such as mutation, crossover, and selection to efficiently converge toward an optimal or near-optimal parameters.  For each `generation' of solutions (i.e., the initial conditions), 
fifteen simulations are run at a time.  Their `fitness' (or goodness-of-fit) is calculated  by comparing the simulated positions and velocities of the LMC and SMC at the current time (the snapshot nearest to when the LMC crosses the Magellanic longitude of 0) with the observed values. 

\begin{equation}
    {\rm fitness} = 1/\sqrt{ \Delta d_{\rm LMC}^2 + \Delta d_{\rm SMC}^2 + \Delta v_{\rm LMC}^2 + \Delta v_{\rm SMC}^2 }
\end{equation}

\noindent
where $\Delta d$ and $\Delta v$ are unnormalized differences in the simulated and observed 3-D positions and space velocities, respectively.
Simulations where the LMC and SMC have merged are downweighted by cutting the fitness value in half
for any simulations where the LMC and SMC are less than 15 kpc apart at the time the LMC crosses $L_{MS}=0$. We tested incorporating comparisons to maps of stellar density, radial velocity, and proper motion in both Magellanic longitude and latitude. However, this did not have much of an effect compared to the central positions and velocities of the galaxies.

The solutions for the next generation are produced with a combination of crossover and mutation. Two solutions are selected with weighted random sampling with a probability proportional to their fitness (`fitter' solutions are more likely to be selected for a pair) and their values averaged. Randomly distributed noise is then added to these values with a Gaussian sigma of 0.5 kpc for the LMC position, 1 km s$^{-1}$ for the LMC velocity, 1 kpc for the SMC position, and 2 km s$^{-1}$ for the SMC velocity. The orientation of the LMC and the galaxy masses are held fixed.
The best-fitting solution of each iteration is always kept to the next iteration.

\begin{figure}
    \centering
    \includegraphics[width=0.95\linewidth]{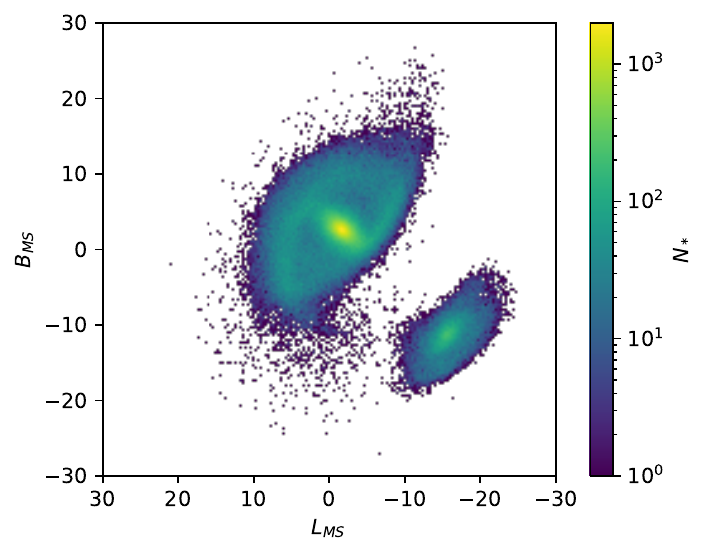}
    \caption{The LMC and SMC at the end of a two-galaxy simulation.}
    \label{fig:2bodyms}
\end{figure}

\subsection{Two-galaxy simulations}
\label{subsec:twogalaxysims}

Initially, we modeled the recent interactions of the MCs using simulations of just the LMC and SMC without the Milky Way, focusing on the recent close interaction. In the end, these simulations failed to provide a good match for our chosen comparison features.  Therefore, we moved on to simulations that also included the MW which are described in the next section.  However, it is instructive to describe the two-galaxy simulations.

\begin{figure*}
    \centering
    \includegraphics[width=0.95\textwidth]{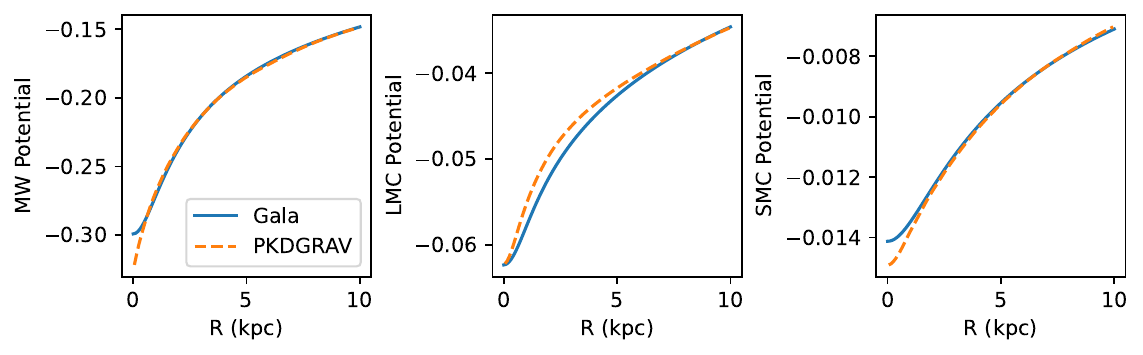}
    \caption{The potentials we use in \gala compared to the potentials of the \pkdgrav\ initial conditions.}
    \label{fig:potentials}
\end{figure*}

\begin{figure*}
    \centering
    \includegraphics[width=0.95\textwidth]{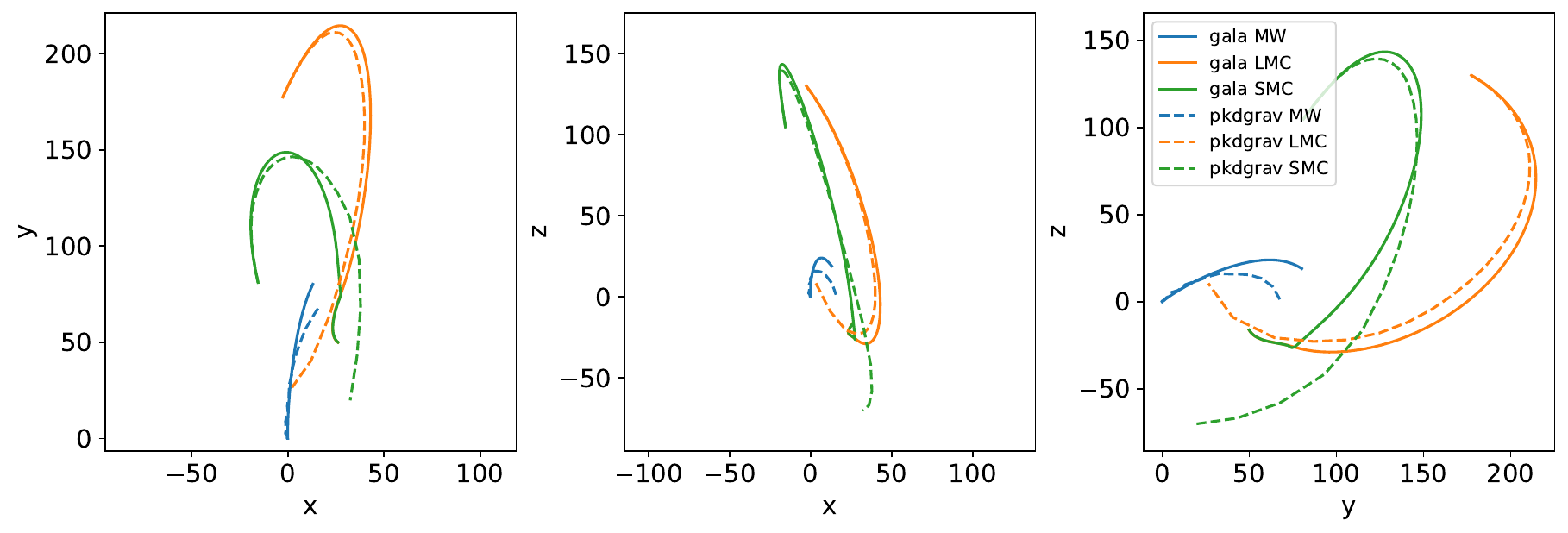}
    \caption{A comparison of orbits with \gala and \pkdgrav\ using the same initial conditions. The \gala orbits are shown as solid lines while the \pkdgrav~orbits are dashed lines. While the orbits are fairly close at the beginning, they diverge and in the \gala run, the LMC and SMC merge. This difference could be due to the fact that with \gala, the galaxies are represented as single particles with fixed potentials, while with \pkdgrav, the shape of the galaxies and thus the shape of their potential wells can change, and stripping causes the dynamical friction and tidal forces to decrease.}
    \label{fig:galapkdcompare}
\end{figure*}

For our simulations involving only the LMC and SMC, we defined a coordinate system that describes the initial locations of the MCs relative to each other as well as where the SMC would first cross the plane of the LMC disc if it were to travel in a straight line. The LMC was placed at the origin with its disc in the $x-y$ plane and defined coordinates as follows: $r$ is the distance from the LMC center to the point, P, where the SMC's path would cross the LMC disc plane if it were moving in a straight line, $d$ is the distance from the SMC center to P, $v$ is the initial speed of the SMC, $\beta$ is the angle from the $x$-axis counterclockwise to P, $\delta$ is the angle counterclockwise of the projection of the SMC's path onto the $x-y$ plane from the $x$-axis, and $\gamma$ is the angle of the SMC's path above or below the $x-y$ plane. The diagrams in \autoref{fig:twogalaxycoords} illustrate the coordinates.

The two galaxies are evolved for 2 Gyr with \pkdgrav.  At the end of each of simulation, the particles are translated so that the center of the simulated LMC is at its current observed coordinates and orientation. In addition, the simulated particles are rotated about the LMC's axis so that the SMC is at its observed orientation angle from our perspective. This allows us to find the locations of the MCs as they would appear in the sky. An example of such a simulation is shown in \autoref{fig:2bodyms}.

We used a genetic algorithm of simulations where we varied the parameters defined above. A total of 4830 simulations were run, however, the final solutions were not satisfactory. In simulations where the SMC ended up closest to its observed position, its velocity was never closer than 75 km s$^{-1}$ from its observed value.
We concluded that the MW's influence on the MCs was an important component that could not be approximated away by this two-galaxy approach.

\begin{figure}
    \centering
    \includegraphics[width=0.95\linewidth]{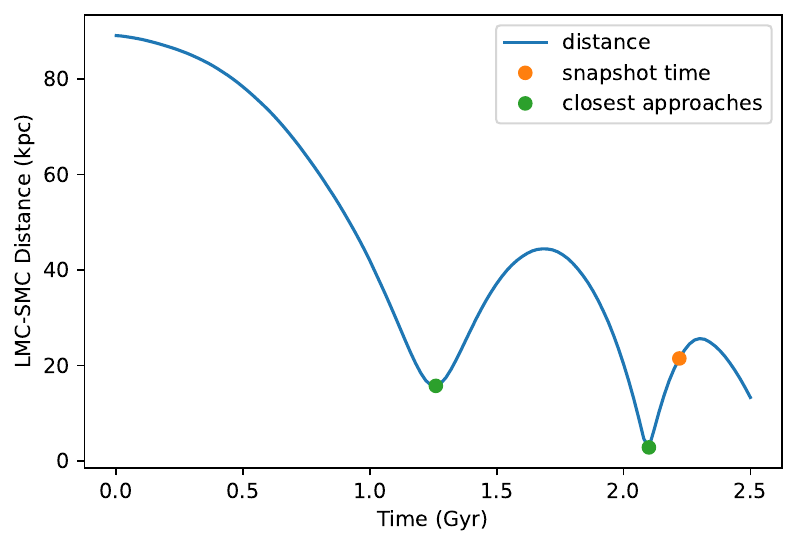}
    \caption{The distance between the LMC and SMC over time for the best match simulation. The green dots are at the closest approaches and the orange dot shows the time for the snapshot when the LMC crosses $L_{\rm MS}=0$\degr.}
    \label{fig:lsmcdist}
\end{figure}

\begin{figure*}
    \centering
    \includegraphics[width=0.95\textwidth]{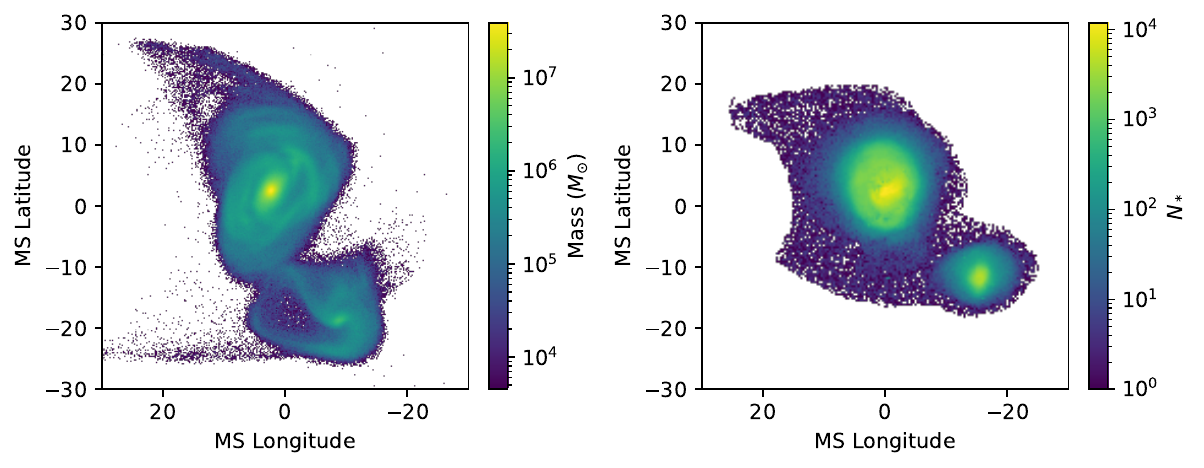}
    \caption{Comparison of density maps in MS coordinates between the best match simulation ({\em left}) and the {\it Gaia} data ({\em right}).}
    \label{fig:simdatamap}
\end{figure*}

\subsection{Three-galaxy simulations}
\label{subsec:threegalaxysims}

Next, we modeled the three-galaxy MW-LMC-SMC system with each component being `live' and modeled as N-body star and dark matter particles.
To find the initial positions and velocities for the model including the MW, we used \gala 
\citep{PriceWhelan2017,gala} run back in time for 2.5 Gyr with MW, LMC, and SMC potentials. The potentials we used for each galaxy in \gala had the same mass for each component as we used with \pkdgrav~above and matched the radial profiles as well. To accomplish this, we used a separate disc and halo component for each galaxy and adjusted their sizes and masses to match the N-body models.
\autoref{fig:potentials} compares radial potentials of our \gala potentials to the \pkdgrav~N-body models. To account for dynamical friction, which is significant in the closely interacting galaxies, we used Equations 8.2 and 8.7 from \citet{Binney2008} to calculate the appropriate approximation to the acceleration. We have dynamical friction from the MW acting on both MCs and dynamical friction from the LMC acting on the SMC.  

As shown in \autoref{fig:galapkdcompare}, the orbits recovered with \gala are close to, but not exactly like, the orbits with \pkdgrav. 
Part of this disparity is due to the fact that the dynamical friction approximation fails when the MCs closely interact.  In an attempt to mitigate this, we increased the dynamical friction on the SMC by the LMC by a factor of seven, which improved the agreement with \pkdgrav~somewhat.
Even with our ad hoc adjustments to the dynamical friction, the differences in the \gala and \pkdgrav~orbits were significant enough that using \gala to determine the best initial conditions was not a viable solution. The increase in dynamical friction, while overall helping the orbits match those found with \pkdgrav, caused us to overestimate the later dynamical friction between the MCs, because the halos do not strip with \gala.
Therefore, we again used a genetic algorithm as with the two-galaxy simulations to find the best initial conditions. We varied the initial positions ($x$, $y$, and $z$) and velocities ($v_x$, $v_y$, and $v_z$) of both MCs.

After running each simulation, the MW is centered and all particles are rotated about the axis of the MW until the LMC is at its observed angle on the x--y plane. Then we find the snapshot in the last 500 Myr where the center of the LMC is the closest to $L_{\rm MS}=0$\degr. This is not necessarily the snapshot with the highest fitness because it does not take the SMC into account but for the purpose of our analysis, we want the LMC to be as close to its observed location as possible. Finally, fitness is calculated based on how well both MC median positions and velocities match the observed values from \autoref{tab:obs}.

In addition, the distances between the MCs are tracked over time including any close interactions they have. The distance versus time plot for the best match simulation is shown in \autoref{fig:lsmcdist}. In the genetic algorithm, the fitness is decreased by half (made worse) for any simulations where the LMC and SMC centers end up less than 15 kpc away from each other at the snapshot with the best LMC position. This prevents the best fit from getting stuck on initial conditions that result in the LMC and SMC merging before the present time.

Once the best initial positions and velocities for the MCs were determined, we also ran a simulation that had just the MW and the LMC starting at the center-of-mass (COM) of the LMC-SMC system. This is used to compare to the three-galaxy simulation to determine which LMC substructures formed as a result of interactions with the SMC.

\begin{figure*}
    \centering
    \includegraphics[width=0.95\textwidth]{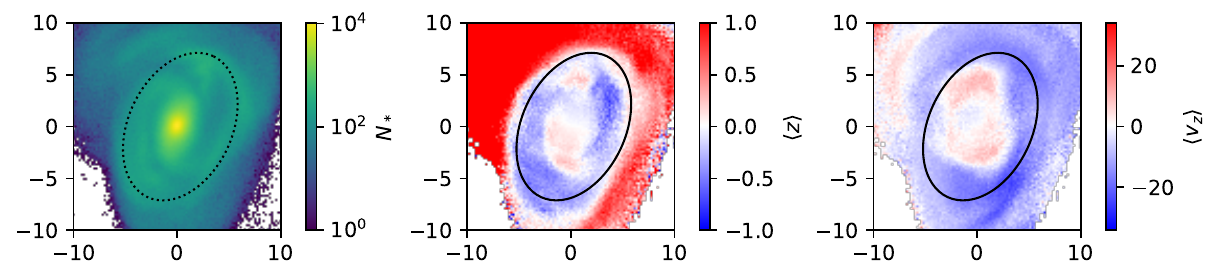}
    \caption{A face-on view of the inner LMC disc, showing the ring and colored by {\em left}: density, {\em center}: mean $z$, and {\em right}: mean $V_z$.}
    \label{fig:lmcring}
\end{figure*}

\section{Results}
\label{sec:results}

\subsection{Comparison to observations}
\label{subsec:resultsobscomparison}

After running $\sim$8000 simulations with the genetic algorithm, we found a best match simulation and ran the same initial conditions with 10$\times$ higher resolution. A comparison between this higher-resolution best match simulation and the data is shown in \autoref{fig:simdatamap}. For the rest of this section, we consider only the best match model.

For the best match simulation, when the LMC crosses $L_{\rm MS}=0$\degr, the LMC position differs from the observed value by 2.8 kpc and its velocity is offset by 14.8 km~s$^{-1}$. The SMC's position differs by 13.6 kpc and its velocity is offset by 13.2 km~s$^{-1}$ from the observed values. The largest discrepancies are the SMC's $z$-coordinate (offset by 12.7 kpc), the LMC's $v_x$ (offset by 10.8 km~s$^{-1}$), and the SMC's $v_y$ (differs by 10.8 km~s$^{-1}$).  The fact that even after our extensive parameter space search the median MC positions and velocities are still offset by these non-negligible amounts shows the difficulty of finding the correct initial conditions.

Even with its shortcomings, the best-match simulation reproduces several observed features including the LMC ring, warp and northern arm. These will be explored in more detail below. Other features, such as the LMC hooks and the SMCNOD, are absent from the simulation.

\begin{figure}
    \centering
    \includegraphics[width=0.95\linewidth]{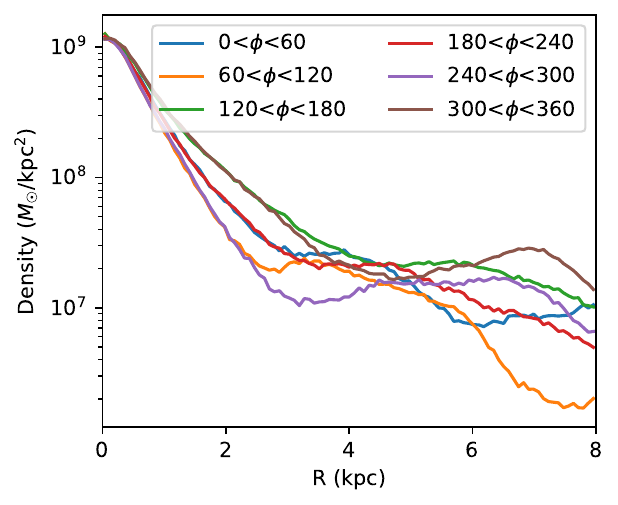}
    \caption{The mass density of the LMC disc as a function of radius in the disc plane for six slices in $\phi$.  The ring overdensity shows up at radii of 4--7 kpc depending on the position angle.}
    \label{fig:ringdensity}
\end{figure}

\subsection{LMC Ring}
\label{subsec:resultslmcring}

In our simulation, there is a high density ring around the center of the LMC that is similar to that described by \citet{Choi2018b} using SMASH photometry of red clump stars. The ring in the simulation has a semi-major axis of approximately 7.5 kpc and a semi-minor axis of approximately 5 kpc, with a position angle of $\sim$30\dgr West of North or $\sim$150\dgr East of North.  The observed ring has a position angle of roughly $\sim$20--30\dgr East of North and a semi-major axis of $\sim$7.5\dgr or $\sim$7 kpc which is similar to what is seen in the simulation. The ring is shown in a map of the LMC in \autoref{fig:lmcring} and the density as a function of radius for different slices of the LMC is shown in \autoref{fig:ringdensity} (similar to the right panel of Figure 1 in \citealt{Choi2018b}).
The ring overdensity appears at radii of $\sim$4--7 kpc depending on azimuth, as expected for an elliptical ring shape.
\citet{Choi2018b} found that the overdensity was 2.5$\times$ higher than the underlying exponential disc at the same radius.  We find that the ring overdensity in the simulation is $\sim$2.8$\times$ higher.

The origin of the ring overdensity has remained unclear and \citet{Choi2018b} postulated two main origin scenarios (1) the evolution of a one-arm spiral, and (2) collisionally-induced by a closer recent interaction with the SMC.
The ring in our simulation is made up of two spiral arms. These arms first became apparent after the first interaction with the SMC, about 1 Gyr ago. After the second interaction 140 Myr ago, these two arms overlap temporarily, creating the ring-shaped overdensity. In future snapshots, the ring expands until the two spiral arms that comprise it separate. Then the current ring disappears, though another smaller ring temporarily forms at the center of the LMC. The LMC ring, therefore, can be added to the mounting evidence of a recent, close encounter of the LMC and SMC.

\begin{figure}
    \centering
    \includegraphics[width=0.95\linewidth]{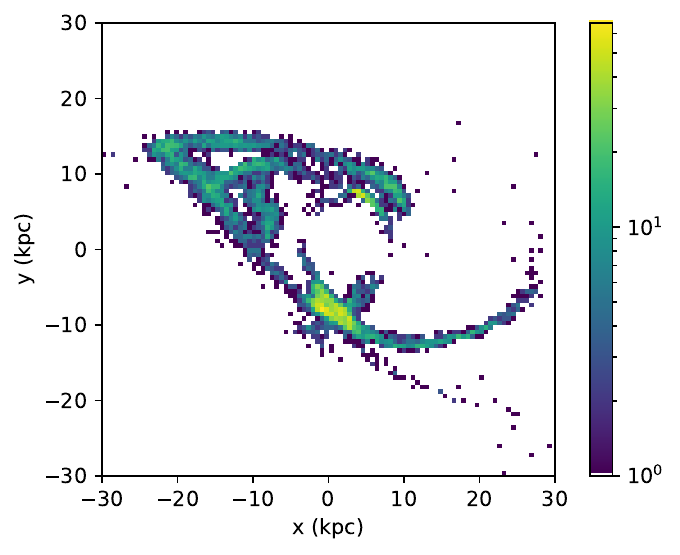}
    \caption{The face-on map of the LMC stars with a deviation value, which is determined by how the velocity components differ from the median for their radius, greater than 20.}
    \label{fig:kinematicfeaturesmap}
\end{figure}

\begin{figure*}
    \centering
    \includegraphics[width=0.95\textwidth]{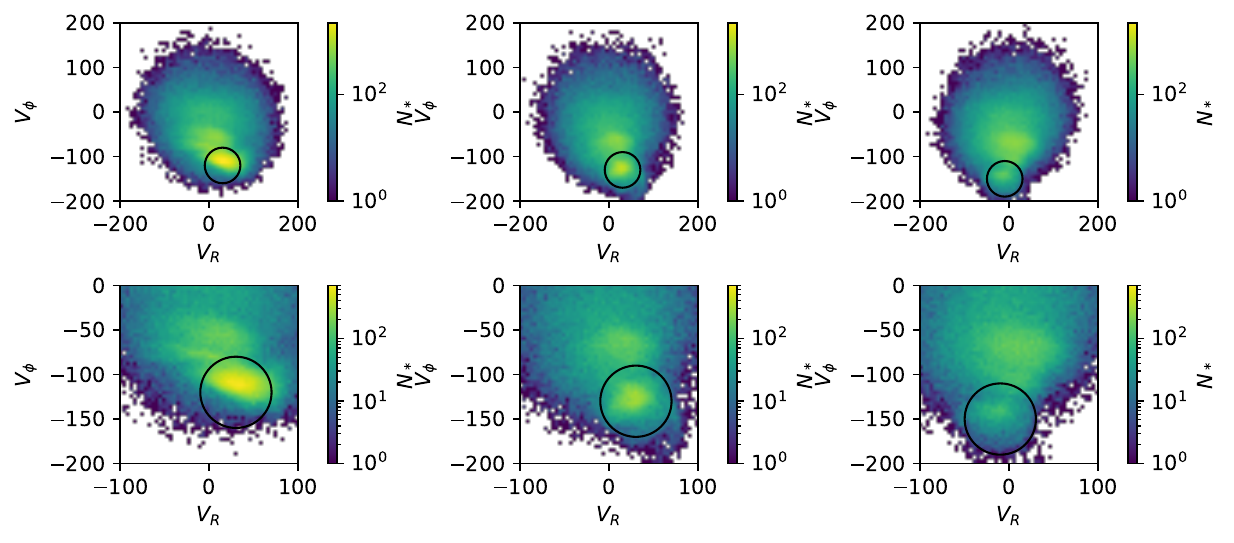}
    \caption{Plots of $v_\phi$ vs $v_R$ for three different angle bins in the LMC ({\em left:} 30\dgre--60\dgre, {\em center:} 60\dgre--90\dgre, {\em right:} 90\dgre--120\dgre). Circles show an overdensity that changes location on the plot for different angles on the LMC. The bottom row shows a zoomed in version. }
    \label{fig:lmcsubstructurevplot}
\end{figure*}

\subsection{LMC kinematic substructure}
\label{subsec:resultslmckinematics}

There are several signatures of LMC kinematic substructure in the simulation. To detect these substructures, we use the following procedure.  The median ($\mu_V$) and robust rms ($\sigma_V$) values of the three velocity components ($V_R$, $V_\phi$, and $V_z$) are measured in 1 kpc radial bins out to a radius of 12 kpc. A `deviation' value is then calculated for each star to determine how deviant its kinematics are from the bulk of the disc at its radius:

\begin{equation}
    {\rm deviation} = \sqrt{ \left( \frac{V_R-\mu_{V_R}}{\sigma_{V_R}} \right)^2 
            + \left( \frac{V_\phi-\mu_{V_\phi}}{\sigma_{V_\phi}} \right)^2
            + \left( \frac{V_z-\mu_{V_z}}{\sigma_{V_z}} \right)^2 }.
\end{equation}

\noindent
Stars in the main disc following normal orbits should have small deviation values, while kinematic substructures should have larger values.
The map of stars that have a deviation over 20 is shown in \autoref{fig:kinematicfeaturesmap}.

Two distinct features are apparent in this map: (1) thin loops in the northeast, and (2) a spiral arm in the southwest that extends several kpc beyond the main disc to the west.
The latter was pulled out by the SMC during their most recent interaction. This tidal arm is remarkably thin, and, therefore, also kinematically `cool' with a velocity dispersion of only $\sim$25 \kmse.
Note that this feature is not visible in the map of the entire simulation in the left panel of \autoref{fig:simdatamap} because it is hidden `behind' the SMC.

\begin{figure}
    \centering
    \includegraphics[width=0.95\linewidth]{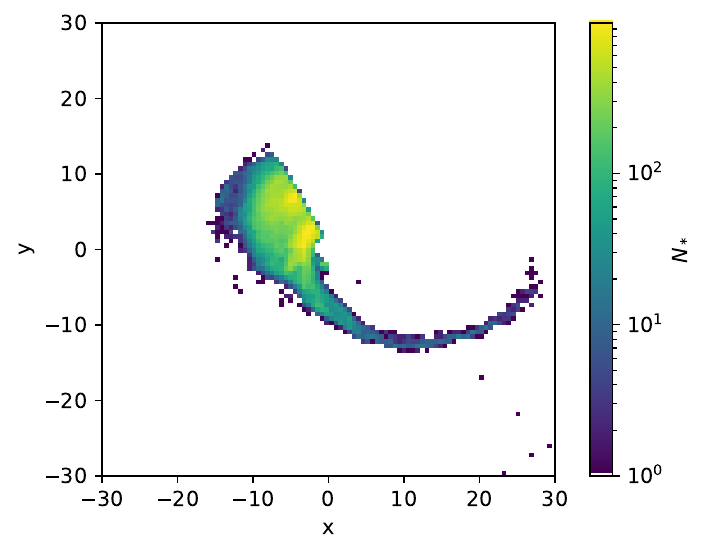}
    \caption{The combined stars of a kinematic feature of the LMC found through two different methods.}
    \label{fig:featurexymap}
\end{figure}

We employ a second technique to investigate kinematic substructure.  
When plotting the different cylindrical velocity components against each other for a limited azimuthal range, there is an angle-dependent overdensity at high $|v_\phi|$.
\autoref{fig:lmcsubstructurevplot} shows this overdensity for three different eastern azimuthal slices.
These two features are part of the same spiral arm as shown in \autoref{fig:featurexymap}. This is also clear from looking at the different components of velocity versus angle $\phi$ in the LMC disc.

While there are several observed LMC substructures in the northeast that could correspond to what we see in the simulation, there is no clear observed counterpart to the western thin tidal arm.  The closest possible match is the southern arm that \citet{Belokurov2019} suggest was tidally induced by the MW and SMC.

\subsection{LMC Northern Stream}
\label{subsec:resultslmcnorthernstream}

Our simulation has two arms in the northern LMC that appear in the map of kinematic features  in \autoref{fig:kinematicfeaturesmap} and can be picked out from plots of the three velocity components as a function of LMC position angle, $\phi$. The northern arm begins at the northwest corner of the LMC and extends to the northeast corner. There is a stream of stars in a nearly straight line from the northeast corner past the southern part of the LMC in \autoref{fig:kinematicfeaturesmap} that are kinematically similar to the northern arm.

\begin{figure*}
    \centering
    \includegraphics[width=0.95\textwidth]{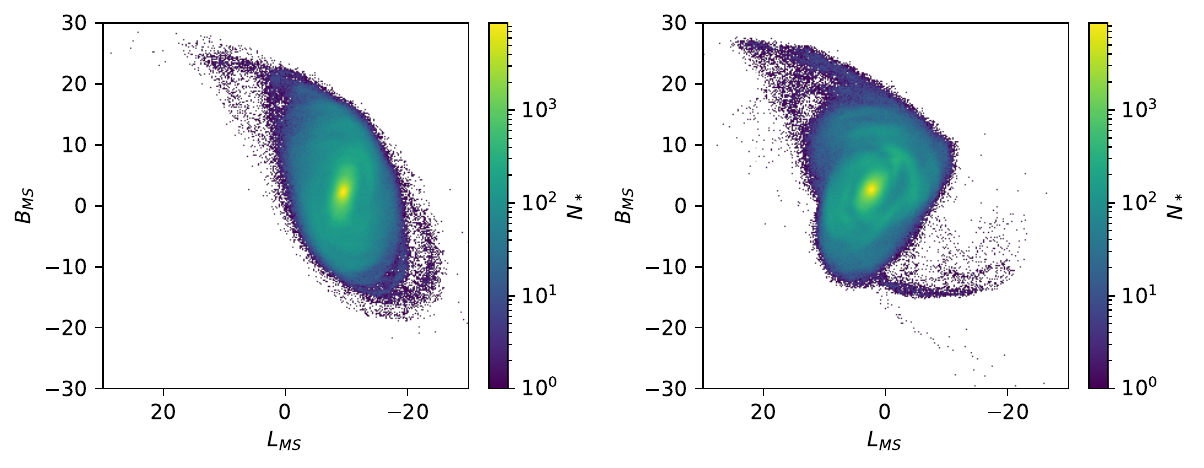}
    \caption{The LMC plotted on Magellenic Stream coordinates in  simulations which exclude ({\em left}) and include ({\em right}) the SMC.}
    \label{fig:lmcwithoutsmc}
\end{figure*}

A northern stream also exists when we run the simulation without the SMC (see \autoref{fig:lmcwithoutsmc}), which suggests that the stream formed at least partially as a result of the interaction between the LMC and the MW. It is likely that the extra influence of the SMC enhances the northern stream.

At the western end of the northern arm, there is a `corner'-like feature where the arm bends at a sharp angle. This is the location where the SMC impacted the LMC during their most recent interaction. The corner does not appear in the simulation without the SMC, as shown in \autoref{fig:lmcwithoutsmc}.

\begin{figure}
    \centering
    \includegraphics[width=0.95\linewidth]{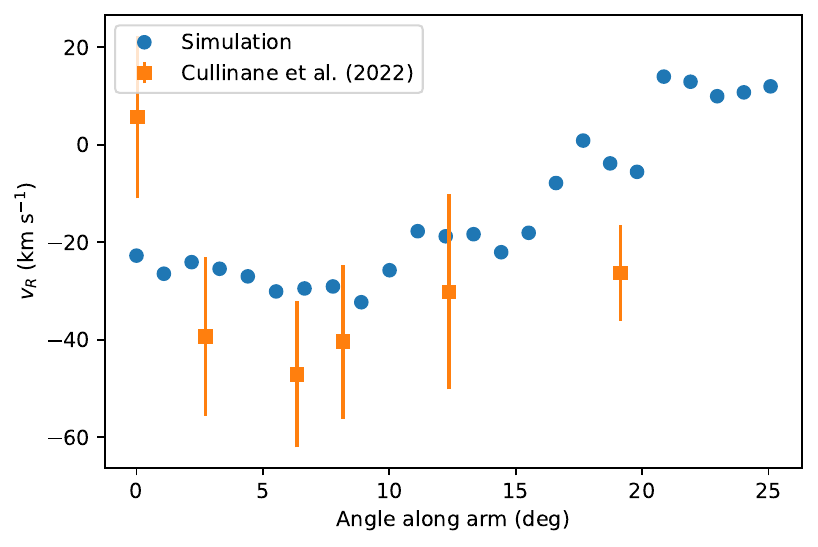}
    \caption{The in-plane radial velocity at different points along the length of our simulated northern arm compared to that measured by \citet{Cullinane2022a}.}
    \label{fig:northarmvr}
\end{figure}

To compare the in-plane radial velocity of our simulated northern arm to observations, we plot the LMC on the sky, and measure the distance in degrees along the arm from where it separates from the rest of the LMC. The comparison is shown in \autoref{fig:northarmvr}. For much of the arm, the magnitude of the velocity is less than observed but the general trend of the velocity being negative and the magnitude increasing at first then decreasing gradually along the arm is similar. Toward the end of the arm in the simulation, the in-plane radial velocity crosses into positive values.

\subsection{LMC disc warp and vertical oscillations}
\label{subsec:resultslmcdiskwarp}

\begin{figure}
    \centering
    \includegraphics[width=0.95\linewidth]{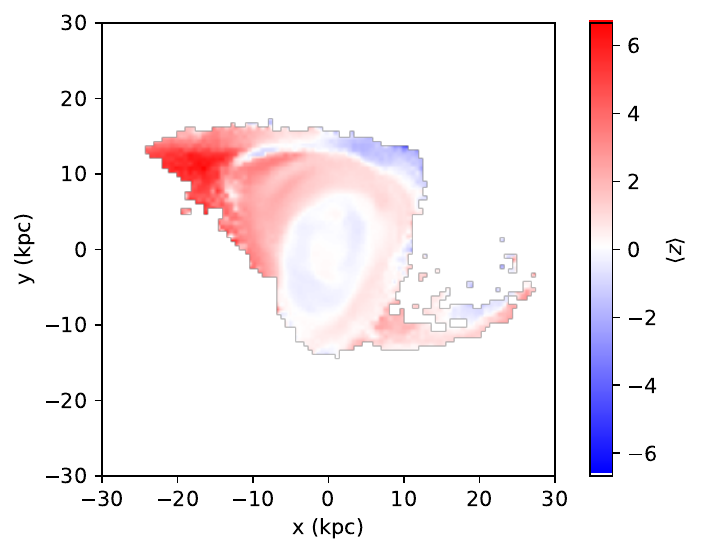}
    \caption{Face-on map of the LMC colored by mean $z$, showing the `ripples' in the disc.}
    \label{fig:lmczmap}
\end{figure}

\begin{figure}
    \centering
    \includegraphics[width=0.48\textwidth]{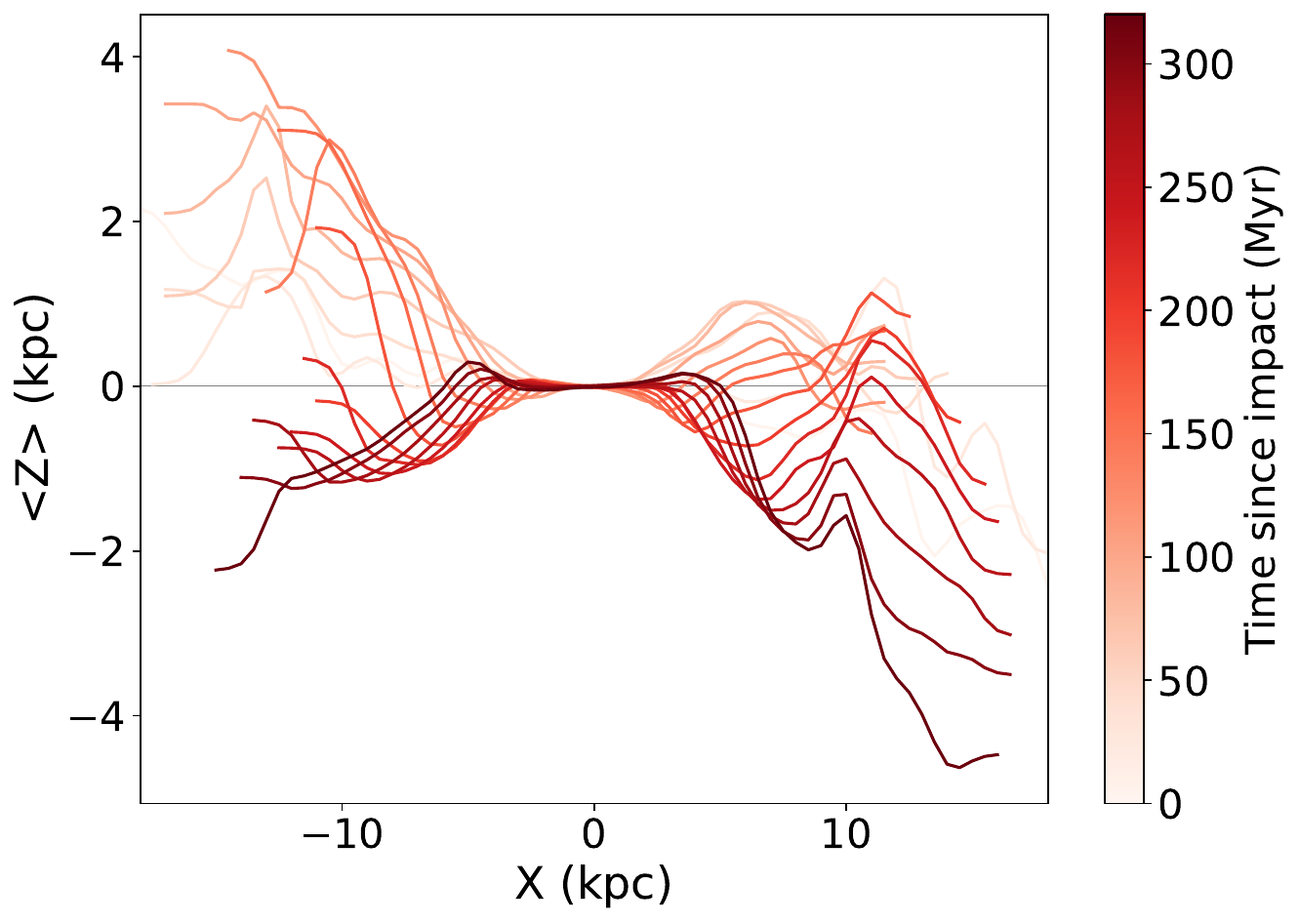}
    \caption{Vertical oscillations in the LMC disc.
    Mean z values versus X in a narrow 2 kpc Y-slit for the 340 Myr after the most recent LMC-SMC collision.  The vertical oscillations are clearly seen in both panels.}
    \label{fig:lmcripples}
\end{figure}

In the high resolution simulation, the LMC disc has a warp with vertical `ripples' in the outskirts
that appear after the most recent interaction with the SMC. At 140 Myr after the interaction, the warp in the LMC outskirts is above the disc plane (closer to the Sun), as defined by the central LMC, while at 300 Myr after the interaction the LMC warp is below the disc plane (farther from Sun). A \meanz~map of the LMC showing the warp and `ripples' 140 Myr after the interaction is displayed in \autoref{fig:lmczmap}.
The `ripples' (or `bending waves') start in the center of the LMC and radiate outward.
\autoref{fig:lmcripples} shows the \meanz~values versus X in a 2 kpc 'slit' in Y as a function of time. This clearly shows the outer disc oscillating vertically with time. Roughly half of a vertical oscillation occurs  in 300 Myr indicating the period is $\sim$600 Myr.

\citet{Choi2018a} detected a southwestern warp in the LMC at a radius of 7\dgr and extending to 4 kpc below the LMC midplane. Later, \citet{Saroon2022} detected the northern component of the warp (towards the northeast) at a radius of 7\dgr and extending 1.2 kpc below the LMC midplane, in the same direction of the southern warp. Tidal interactions generally create S-shaped warps \citep[e.g.,][]{Toomre1972},
therefore, the detected U-shaped warp was somewhat unexpected. However, the U-shaped warp could be an effect of the bending waves that we detect in our simulation that are produced in a direct collision of the LMC and SMC.
The strong oscillations start around a radius of 7--8 kpc in the simulation which is at a similar radius to the observed warps.

\subsection{SMC Expansion and Rotation}
\label{subsec:resultssmcexpansion}

\begin{figure}
    \centering
    \includegraphics[width=\linewidth]{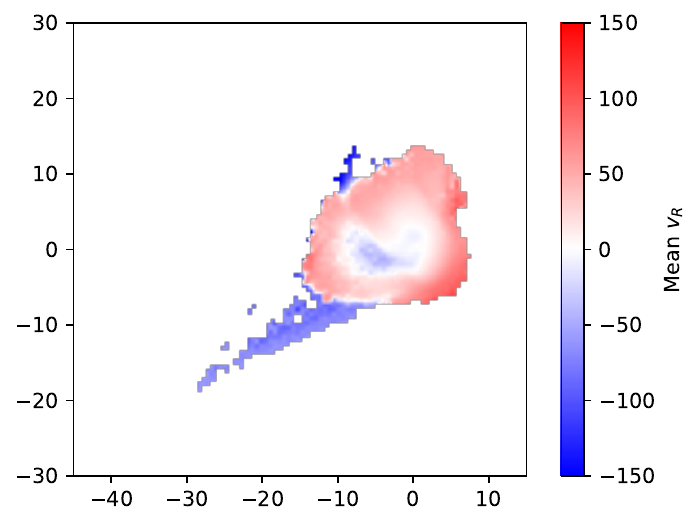}
    \caption{Face-on map of the SMC colored by mean $v_R$.}
    \label{fig:smcexpansion}
\end{figure}

\begin{figure*}
    \centering
    \includegraphics[width=0.47\linewidth]{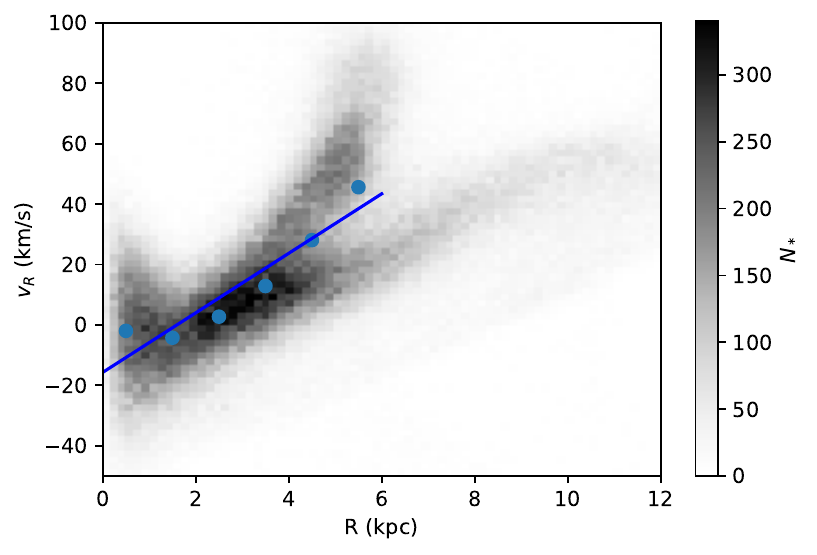}
    \includegraphics[width=0.47\linewidth]{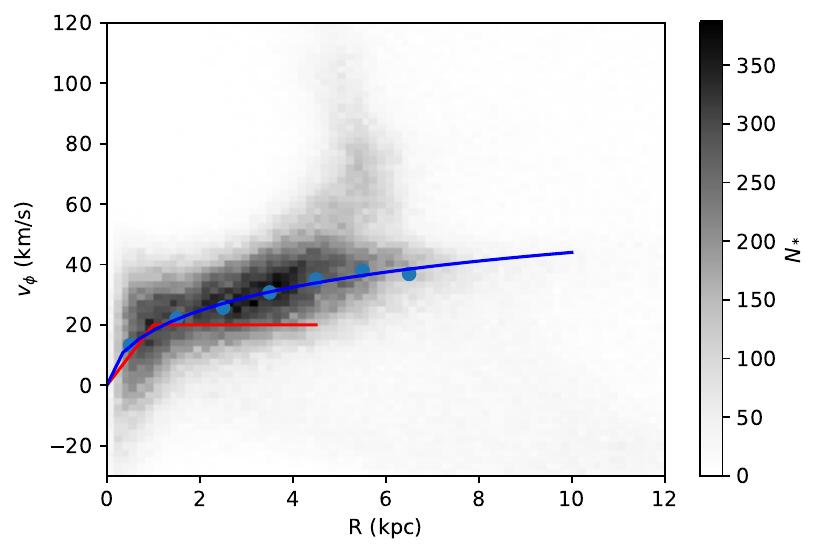}
    \caption{Kinematics of SMC star particles. (Left) $v_R$ versus $R$. The filled circles show the median velocity value for each of the first six 1 kpc radial bins.
    A linear fit of 9.86 km s$^{-1}$ kpc$^{-1}$ is shown by the blue line.
    (Right) $v_\phi$ vs $R$. The median for each 1 kpc radial bin is again shown. A rotation curve matching the median values is shown in blue.  The \citet{Zivick2021} results are in red.}
    \label{fig:smcrvr}
\end{figure*}

\begin{figure}
    \centering
    \includegraphics[width=0.95\linewidth]{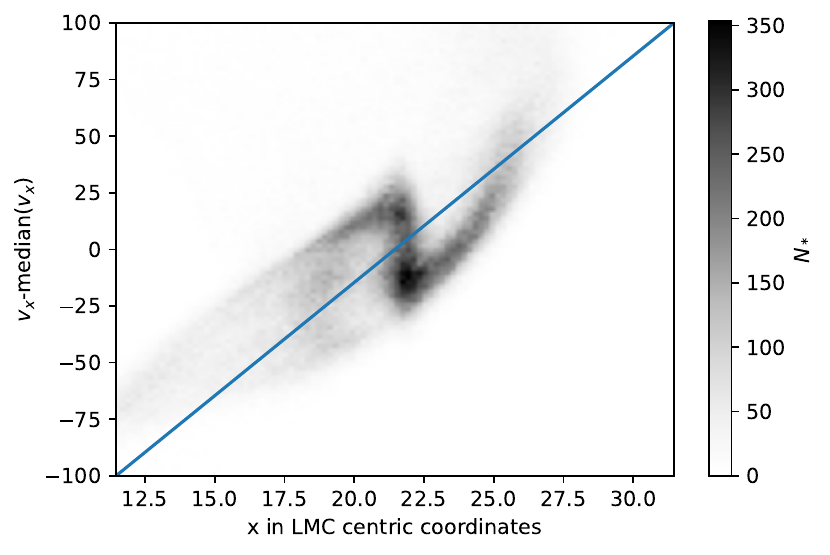}
    \caption{The tidal expansion of the SMC. The coordinate system here is centered on the LMC and the SMC is on the x-axis, so $v_x$ is the component of the SMC's velocity toward or away from the LMC (negative is towards the LMC). The blue line shows the tidal expansion of 10 km s$^{-1}$ kpc$^{-1}$ found by \citet{Zivick2021}.}
    \label{fig:tidalexpansion}
\end{figure}

In our Model A high-resolution simulation, the SMC radially expands after the most recent interaction with the LMC. The outward radial velocity of the stars increases with radius, as can be seen in the $v_R$ map (\autoref{fig:smcexpansion}), while the linear radial dependence can be seen more clearly in the left panel of \autoref{fig:smcrvr}.
We find that the slope of $v_R$ as a function of $R$ to be 9.86~km~s$^{-1}$~kpc$^{-1}$ extending to a radius of 6 kpc. The kinematic analysis of \citet{Zivick2021} of the SMC {\it Gaia} DR2 proper motions (to a radius of $\sim$4.5\degr) found a tidal expansion term of $10\pm1$~km~s$^{-1}$~kpc$^{-1}$ in the direction of the LMC. \autoref{fig:tidalexpansion} shows the tidal expansion of the SMC in the direction of the LMC in our simulation. The Zivick et al.\ result is indicated by a blue line indicating a good overall agreement.

Before the LMC-SMC close interactions, the SMC's rotation velocity is 60.5 km s$^{-1}$ and reaches a maximum at a radius of $\sim$3 kpc in our simulation.  After the most recent LMC-SMC impact, the SMC's rotation is diminished, but there remains a clear `leftover' rotation in the simulation. The rotation velocity increases with radius until it reaches a maximum of $\sim$38 \kms at a radius of $\sim$6 kpc (right panel of \autoref{fig:smcrvr}). For stars at a radius around 1 kpc, the median is about 18 \kmse. This is within the range of 15--25 \kms for rotation at 1 kpc found by \citet{Zivick2021}.
The Zivick et al.\ kinematic analysis determined a maximum rotation velocity of 20 \kms reaching the maximum at 1 kpc.  This is lower than the value we find in the simulation. The $v/\sigma$ for the SMC remains fairly constant at $\sim$3.5 before the first interaction, at which point it drops to $\sim$2.5. At the most recent interaction at 2.06 Gyr, it drops again to $\sim$1.0 and continues to decrease sharply from there. In the `current' snapshot, $v/\sigma=0.79$. This is slightly more than the observed value of $v/\sigma \leq 0.6$ \citep{Harris2006}, but it reaches this value about 60 Myr after the `current' snapshot.

\subsection{SMC Kinematic Substructure}
\label{subsec:resultssmckinematicsubstructure}

\begin{figure}
    \centering
    \includegraphics[width=0.95\linewidth]{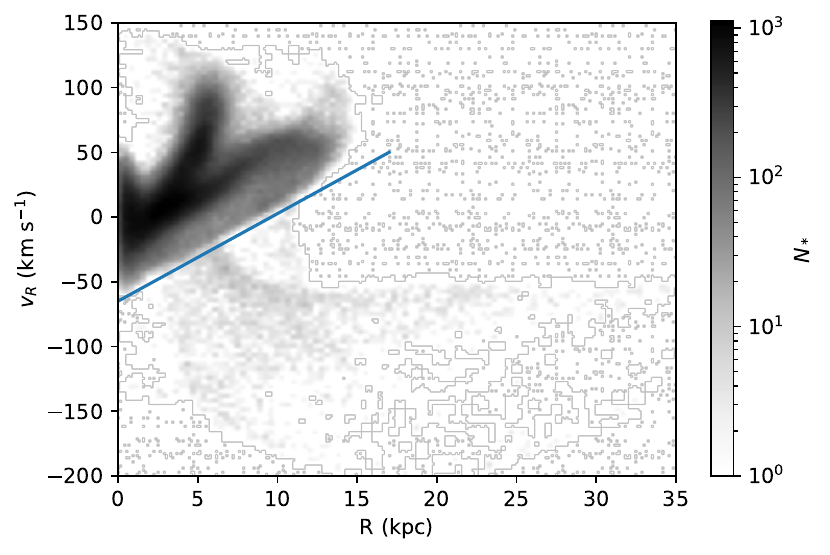}
    \caption{SMC $v_R$ vs. $R$ with different axis limits than \autoref{fig:smcrvr} to show stars with higher radii and negative $v_R$. The blue line is at the cutoff between the main SMC disc and a kinematic feature.}
    \label{fig:smcrvr2}
\end{figure}

\begin{figure*}
    \centering
    \includegraphics[width=0.95\textwidth]{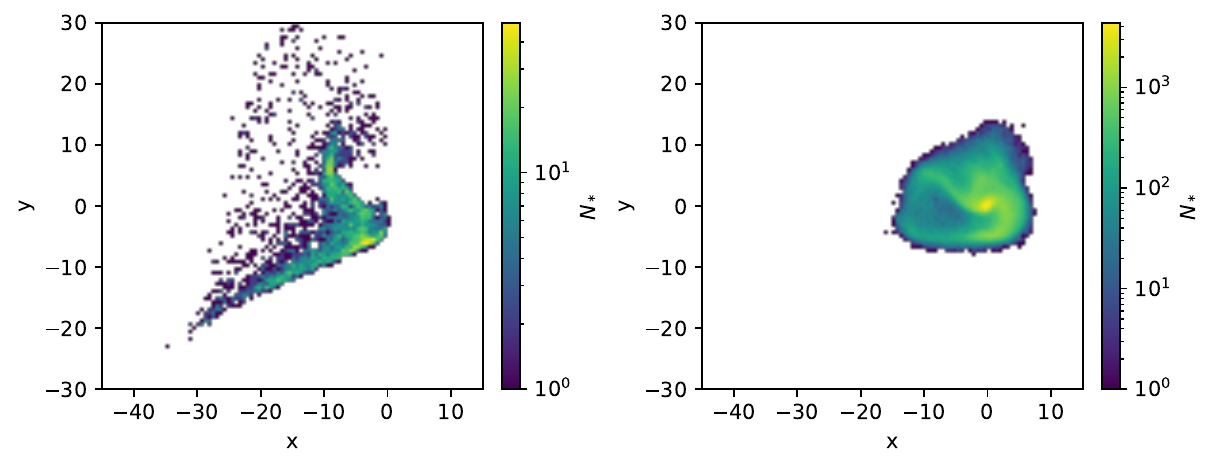}
    \caption{The face-on SMC in the SMC-centric coordinate system. {\em Left:} The stars below the cut in \autoref{fig:smcrvr2}. {\em Right:} The stars above that cut.}
    \label{fig:smcfeaturexy}
\end{figure*}

In the SMC, there is a substructure that can be seen in the plot of $v_R$ vs.\ $R$, shown in \autoref{fig:smcrvr2}. As shown in \autoref{fig:smcfeaturexy}, the stars below the cut in \autoref{fig:smcrvr2} are mostly in two arms that start near the center and extend out to a larger radius than the main SMC. The stars in the arm to the south were pulled away from the SMC during the collision with the LMC, while the one to the east mostly remains within the SMC disc. The stars in the southern arm have a similar velocity in the $x$ direction ($\sim$110 \kmse) and in the simulation continue past the main SMC after the current snapshot. While there is observed stellar substructure called the SMC Southern Overdensity \citep[SMCSOD;][]{Massana2024} it is as of yet unclear if these are the same structures. This feature in our simulation does not appear to be kinematically similar to the observed SMCSOD. The stars in the observed SMCSOD tend have a proper motion $\mu_B$ equal to or greater than the main SMC \citep{Massana2024}, while the $\mu_B$ for the stars in the simulated feature is lower than for the main SMC.

\subsection{SMC Stellar Radial Density Profile}
\label{subsec:resultssmcdensityprofile}
\begin{figure*}
    \centering
    \includegraphics[width=0.47\linewidth]{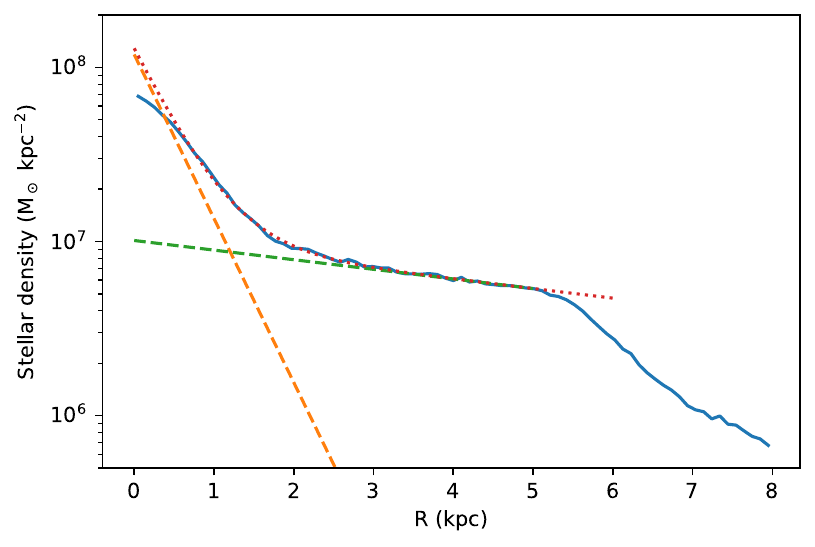}
    \includegraphics[width=0.47\linewidth]{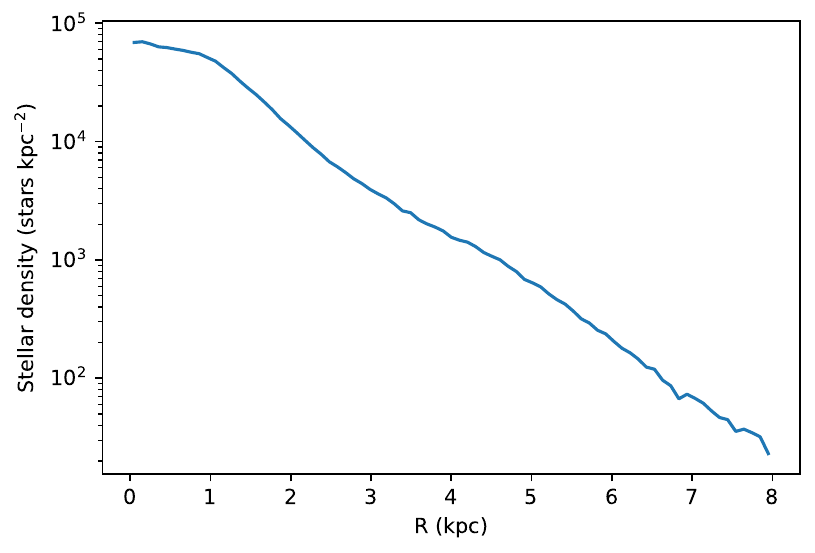}
    \caption{The radial density of the SMC ({\em left}) in the simulation with two exponential discs (dashed lines) and their sum (dotted line) ({\em right}) from {\it Gaia} data.
    }
    \label{fig:smcradialdensity}
\end{figure*}

The simulated SMC disc density can be modeled as two exponentials out to a radius of $5.5$ kpc, one with scale length $0.46$ kpc and the other with scale length $7.87$ kpc. The first models the central SMC `core', 
while the second accounts for the more extended main body and tidal arms. At larger radii, the density of the SMC breaks and drops off after $5.5$ kpc,
as shown in \autoref{fig:smcradialdensity}. This drop in density is likely not due to the truncation of the SMC disk, since the overdensity is moving outward from the center of the SMC at the time of this snapshot. We compare the simulation profile to the {\it Gaia} SMC red giant branch starcount data (right panel of \autoref{fig:smcradialdensity}), which show that the SMC can be modeled fairly well with a single exponential profile for most of its extent,
with a slight overdensity at a radius of $\sim$4--5 kpc.
Observations show the SMC has a more elliptically-shaped core in the inner 1--2\dgr but is more circularly shaped at larger radii \citep{Luri2021}. These could correspond to the two components seen in the simulation.
In the periphery, \citet{Nidever2011} and \citet{Pieres2017} both found extended SMC stellar components beyond $\sim$7\dgr which in the north corresponds to the SMCNOD. There is no clear counterpart to the SMCNOD in our simulation.

\subsection{SMC Distance Bimodality}
\label{subsec:resultssmcdistancebimodality}

\begin{figure}
    \centering
    \includegraphics[width=0.95\linewidth]{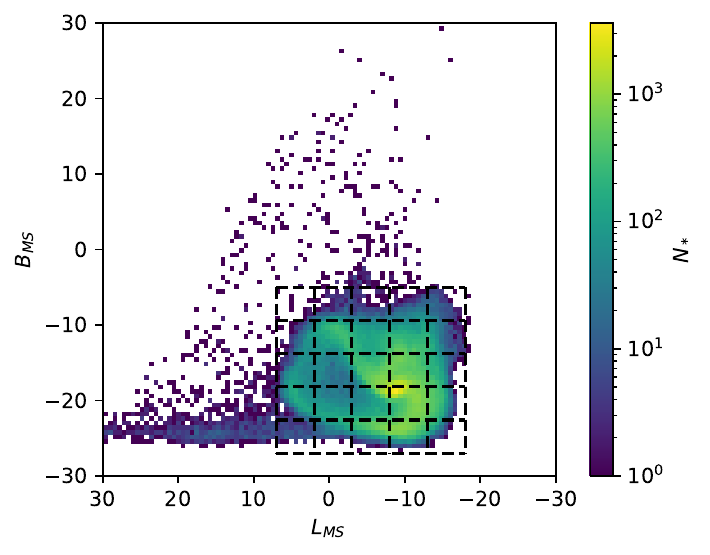}
    \caption{Map of simulated SMC stars on Magellanic Stream coordinates with dashed lines dividing the SMC into different regions.}
    \label{fig:smcdistboxes}
\end{figure}

\begin{figure*}
    \centering
    \includegraphics[width=0.95\textwidth]{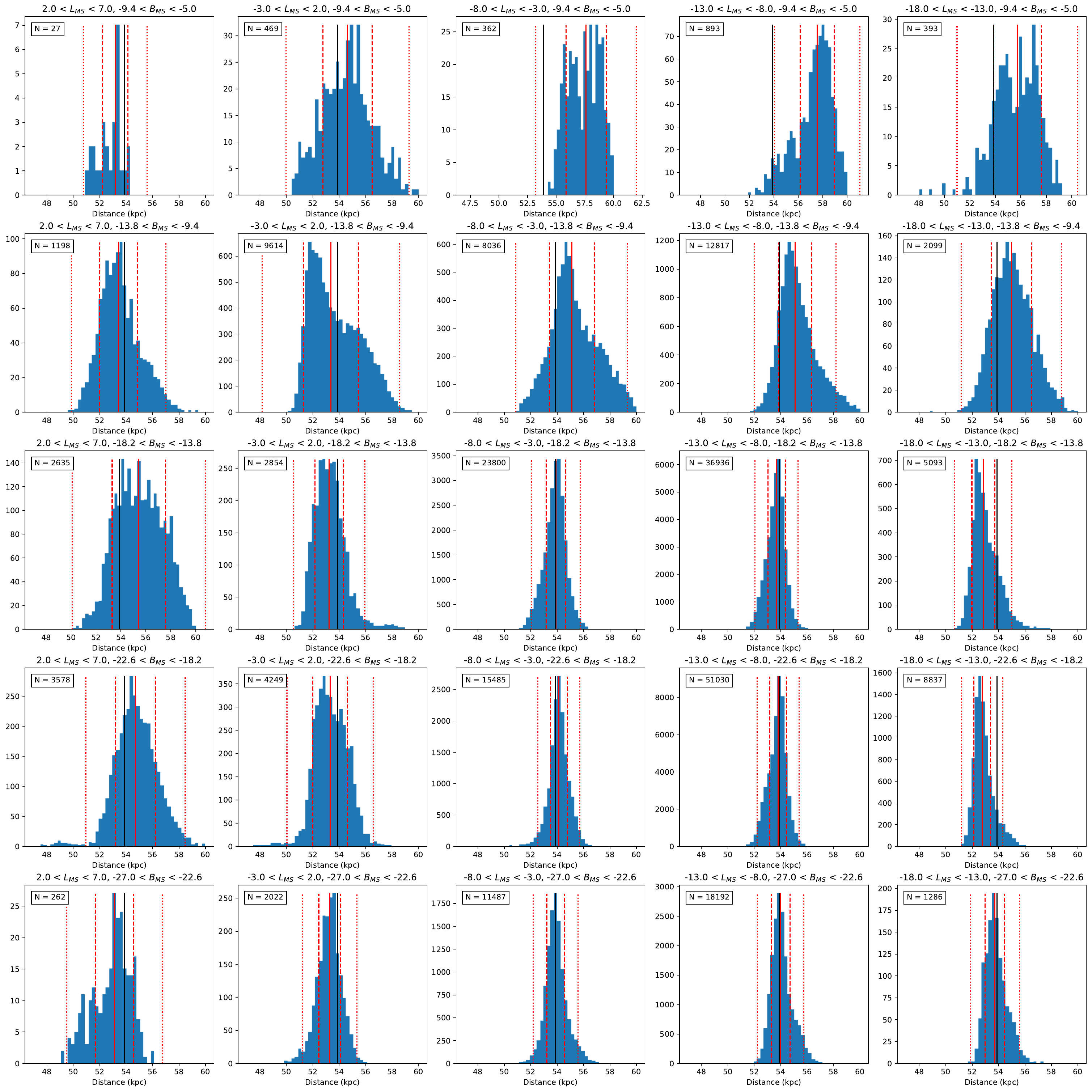}
    \caption{Histograms of distances to simulated SMC stars. Each histogram is of stars in the box in the same grid location as in \autoref{fig:smcdistboxes}. The solid red line is the median distance, the dashed lines are one median absolute deviation (MAD) away from the median, and the dotted lines are 2.5 MADs away from the median. The black line is the median distance of all the SMC stars.}
    \label{fig:smcdisthists}
\end{figure*}

The SMC exhibits a distance bimodality in the east, towards the direction of the LMC, which is clearly seen in the CMD distribution of red clump stars \citep[e.g.,][]{Hatzidimitriou1989,Nidever2013}. The nearer group of stars, which are $\sim$10 kpc closer to the Sun than the main SMC, were likely separated from the SMC by tidal interactions with the LMC \citep{Almeida2024}. With the simulated SMC, we divide the SMC into a spatial 5$\times$5 grid (see \autoref{fig:smcdistboxes}) and show their distance distributions in \autoref{fig:smcdisthists}.
The region where the distance bimodality is observed lies at the top of the figures at higher $B_{\rm MS}$. The central and western (bottom panels) regions show a fairly narrow single-peaked distance profile consistent with observations, while the eastern regions exhibit a clear bimodality resembling that observed.
The bimodality appears after the most recent interaction between the MCs.

\begin{figure*}
    \centering
    \includegraphics[width=0.95\textwidth]{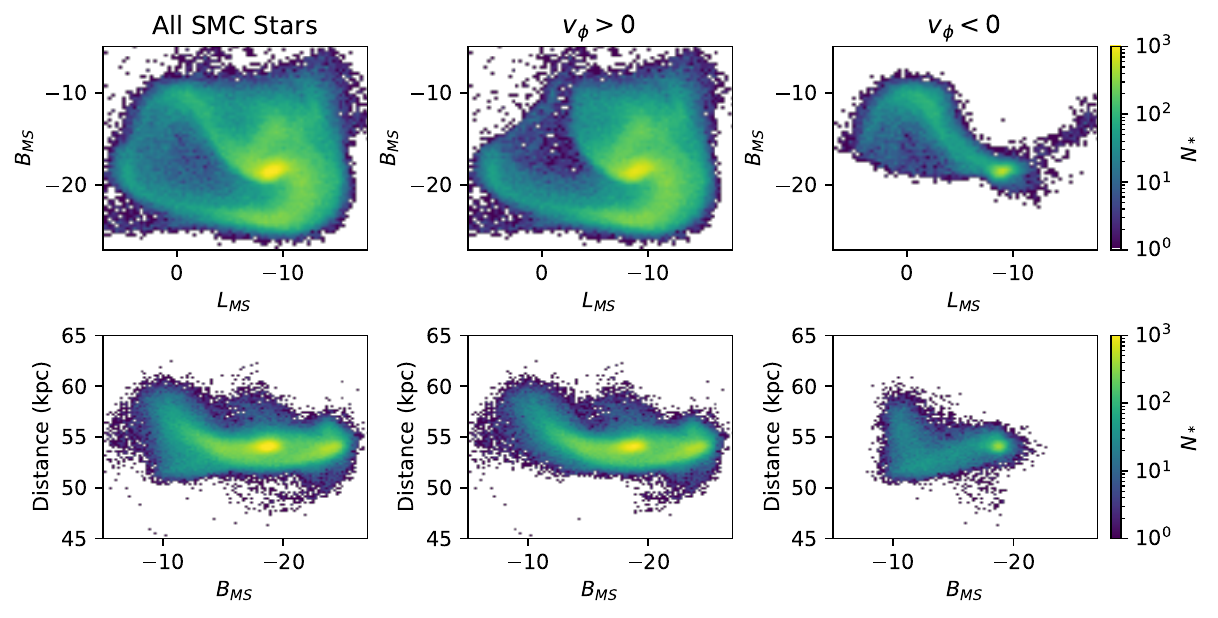}
    \caption{Density map on the sky and distance vs. $B_{MS}$ for simulated SMC stars. The left panels include all SMC stars, the center panels include only the stars where $v_\phi>0$, and the right panels include only the stars where $v_\phi<0$.}
    \label{fig:smcretrograde}
\end{figure*}

The eastern SMC is composed of two components. These are easily separated in azimuthal velocity in the SMC-centric frame into a prograde and retrograde component. \autoref{fig:smcretrograde} shows the spatial density of all stars and the two components in the top panels, while distance vs.\ $B_{\rm MS}$ is shown in the bottom panels. The prograde component is more distant and has a spiral arm-like shape. In contrast, the retrograde component is closer and has the shape of a tidal arm extending from the central SMC. In addition, the stars that make up the retrograde component all come from the inner SMC ($R$$<$5 kpc).
This is consistent with the results of \citet{Almeida2024} who found that the closer SMC component is tidally stripped from the inner SMC.

\subsection{Multiple LMC Initial Conditions}
\label{subsec:resultsinitialconditions}

\begin{figure*}
    \centering
    \includegraphics[width=0.95\textwidth]{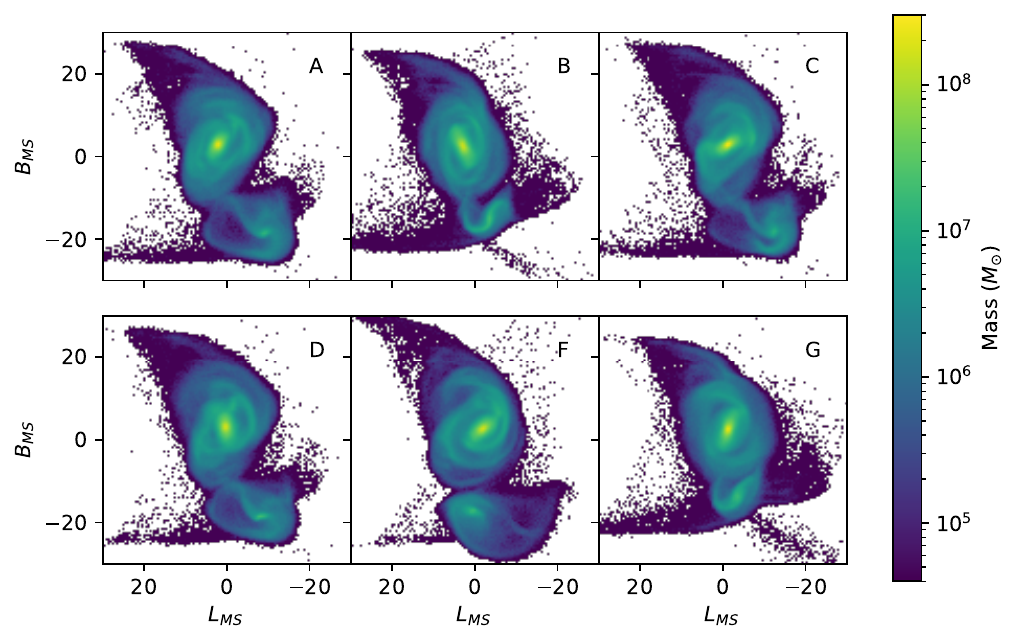}
    \caption{The maps of the different simulations at the `current' snapshot for each model. {\em Top row:} models A, B, and C. {\em Bottom row:} models D, F, and G.}
    \label{fig:lmcmodelsmaps}
\end{figure*}

We also ran the simulation with the same starting positions and velocities but with different LMC models as described in \autoref{sec:simulations}. The maps of the different models at the `current' snapshot are shown in \autoref{fig:lmcmodelsmaps}.
In all models, there are two interactions between the MCs that occur at about the same times. Every model also ends up with a northern stream, two arms forming a high density ring in the LMC disc, and an arm pulled out from the LMC like the feature described in \autoref{subsec:resultslmckinematics}.
Therefore, we conclude that these are robust features of the models.

Other things, such as the time at which the LMC crosses the MS longitude zero and the position of the SMC, differ between models. In both models where we change the thickness of the LMC disc (B and C) as well as the one with a more diffuse dark matter halo (G), the LMC crosses MS longitude zero earlier than in the fiducial simulation. The time difference is 20 Myr in model C, 40 Myr in model G, and 80 Myr in model B. Since the timing of when the LMC crosses $L_{MS}=0$ is how we define the `current' snapshot, it might be significant to our other results that these simulations did not run as long as the others, which had the LMC cross zero at the same time as for model A. In two of these simulations (models B and G), the location of the SMC is more directly south of the LMC than for other models. The SMC is also in a similar location south of the LMC in model F. In the others, the SMC is in a similar location in the sky as for the fiducial simulation.

The direction and amplitude of the warp at the `current' time differ substantially across the models.
Models A, C, and D all exhibit a positive warp (toward the Sun), with model D evolving in a way that closely mirrors model A. In contrast, models B and G have a negative warp (away from the Sun). Model F shows no coherent warp in either direction; instead, it develops only small-scale ripples.
The radial extent of these ripples also varies with disc thickness. In model B, which uses the thicker disc, the ripples appear only in the outer LMC (beyond $\sim$8 kpc). In model C, with the thinner disc, they begin much farther in, at roughly 5 kpc. Finally, the LMC disc for model G bends in both directions at different azimuths rather than forming a uniform, global warp.

Overall, the parameter that had the most effect on the simulation when changed was the concentration of the dark matter halo. The range of parameter values we looked at left much uncertainty in the characteristics of the LMC.

\section{Discussion}
\label{sec:discussion}

While it is possible to `successfully' reproduce observational results with simulations, such as the Sagittarius stellar stream by \citet{Law2010} and the M31 great southern stream by \citet{Fardal2013}, it is generally challenging to match simulations to data.
One exception is the Made-to-Measure technique \citep{Syer1996}, which fits a system to observations by varying particle weights within a prescribed potential --- an approach distinct from fully self-consistent N-body simulations and less suited to strongly non-equilibrium systems like the Magellanic Clouds.
The Magellanic System is notoriously difficult to model due to the multiple strong interactions and the complexity of the system. Even with more than 8000 simulations, our genetic algorithm search of the parameter space found an approximate match to the central MC positions and velocities. Even so, there are a number of observational stellar features that are reproduced in our best-match simulation.

The LMC ring identified and characterized by \citet{Choi2018b} is clearly reproduced in our simulation. While the origin of the ring remained unclear, we find that it is collisionally-induced by the recent direct SMC-LMC collision. The ring is made up of two overlapping arms in the LMC, and most of the stars in the ring are moving outward with a median $v_R$ of 12.7 \kmse. This ring is present in all the simulations we ran with different LMC parameters, so it is robust.

In addition, our simulation reproduces the northern LMC stream which is a prominent observational substructure in the LMC outskirts. While \citet{Belokurov2019} were able to generate a northern stream with a simulation of only the LMC and MW, we find that the presence of the SMC significantly enhances the northern stream and modifies its shape. The northwestern edge of the stream becomes substantially kinked and the western edge of the LMC disc quite linear similar to what is observed.

The recent direct collision of the LMC and SMC also substantially affected the structure and kinematics of the SMC. The simulated SMC stars have a significant positive radial velocity component especially beyond a radius of $\sim$2 kpc. This is similar to the tidal expansion observed by \citet{Zivick2018}. Even with the dramatic disturbance to the SMC's structure, it shows residual rotation -- although at levels $\sim$2$\times$ that seen by Zivick et al.\ in the {\it Gaia} data.

The most pronounced SMC substructure is the distance bimodality on the eastern side. This is also found in our simulation where the eastern side of the SMC is composed of two components. The prograde component is more distant and part of the main SMC structure, whereas the retrograde component is closer and tidally pulled from the SMC center. This confirms the findings of \citet{Almeida2024} which showed that the metallicity distribution function of the closer eastern component is very similar to the central SMC, while the more distant eastern component is like the western SMC periphery.

A striking phenomenon in our high resolution interaction simulation is the vertical oscillations of the outer LMC disc. This is a direct result of the recent collision of the two galaxies. These oscillations appear in most of the simulations where we vary LMC parameters, but are less pronounced in the one with more concentrated dark matter halo. While similar vertical oscillations have been observed in the Milky Way, which could be due to the influence of Sagittarius dwarf spheroidal and likely the LMC \citep[e.g.,][]{Gomez2016,Laporte2018}, such oscillations have not yet been detected in the LMC. 
We predict that such vertical oscillations should be observable in the LMC, especially at larger radii, if the LMC and SMC had a direct collision.
The U-shaped warp detected by \citet{Choi2018a} and \citet{Saroon2022} could be the inner edge of the vertical oscillations. We note that \citet{Oden2025} recently discovered that the warp is indeed U-shaped and extends to a radius of $\sim$15 kpc.

Our simulation does not reproduce the LMC southern hooks or the SMCNOD feature. It is quite possible that these were created by interactions of the MCs before 2.5 Gyr ago. In future work, we plan to start the simulations at earlier times to capture some of these events.

\section{Conclusions}
\label{sec:conclusions}

We use a suite of N-body simulations of the LMC, SMC and MW system to reproduce observed stellar features in the Magellanic System.
Using a genetic algorithm, we ran over 8000 simulations of the three galaxies until we found initial conditions that resulted in the MCs being close to their observed locations after two close interactions in the past 2.5 Gyr.

Our main findings are as follows:
\begin{itemize}
    \item There are two close passages between the MCs in the past 2.5 Gyr. In our best simulation, they are about 940 Myr and 140 Myr before the LMC reaches its current position.
    \item The simulation reproduces the observed LMC ring which is a temporary structure of two collisionally-induced LMC spiral arms by the recent close SMC-LMC collision which are moving radially outward.
    \item The LMC northern stream is reproduced by a combination of MW and SMC tidal forces on the LMC. The influence of the SMC enhances the feature and creates a unique kinked shape similar to what is observed.
    \item The simulated SMC exhibits radial expansion after the recent collision which is similar to the observed tidal expansion. In addition, there is residual rotation as also observed, but at a level $\sim$2$\times$ higher.
    \item The observed distance bimodality in the eastern SMC is reproduced. This is due to two eastern components, one of which is a retrograde tidal arm pulled from the central SMC and more nearby. This is consistent with recent observations of SMC abundances.
    \item Kinematic substructures are observed in both the LMC and SMC that do not have any clear observational counterparts.
    \item The warp in the outer LMC shows pronounced vertical oscillations from the last SMC encounter. The U-shaped warp is consistent with observational results of the southern and northern warps in the LMC periphery.
    \item Changing the thickness of the LMC disc affects the size and direction of the warp. In our simulation with the higher density LMC halo, only a small warp appears, while in the one with the lower density halo, the warp is S-shaped rather than U-shaped.
\end{itemize}

\section*{Acknowledgements}

B.R.G. and D.L.N. acknowledge support from National Science Foundation grants AST 1908331 and 2408159. 

Computational efforts were performed on the Tempest High Performance Computing System, operated and supported by University Information Technology Research Cyberinfrastructure (RRID:SCR\_026229) at Montana State University. Some simulations were also run on the Flatiron Institute's Rusty computing cluster, which is supported by the Simon's Foundation. Additional early simulations were run at the High Performance Computer Facility
of the University of Lancashire.

Portions of the text were refined with the assistance of OpenAI’s ChatGPT (GPT-5), which was used to improve clarity, grammar, and style.


\section*{Data Availability}

Our fiducial simulation is available at https://doi.org/10.5061/dryad.1vhhmgr82



\bibliographystyle{mnras}
\bibliography{ref-og}



\appendix


\bsp	
\label{lastpage}
\end{document}